\documentclass[11pt, a4paper]{article}

\usepackage{jheppub}
\usepackage{amsfonts}
\usepackage{amsbsy}
\usepackage{amsthm}
\usepackage{epstopdf}

\newtheorem{theorem}{Theorem}[section]

\newcommand{\fl}[1]{ \lfloor #1 \rfloor }

\def\Z{\mathbb{Z}}
\def\C{\mathbb{C}}
\def\R{\mathbb{R}}
\def\CF{\mathcal{F}}
\def\HZ{H_2(S,\Z)}
\def\lat{\Gamma^{3,19}}
\def\lattice{\Gamma^{3,19}}
\def\Tr{\mathop{\mathrm{Tr}}}
\def\nb{\overline{N}}

\def\M{\mathcal{M}}

\def\gcd{\mathrm{gcd}}
\def\F{\mathbb{F}_2}
\def\P{\mathbb{P}}
\def\L{\mathcal{L}}
\def\N{\mathcal{N}}
\def\det{\mathrm{det}}

\def\fwvec{\frac{1}{2}\sum_a N_a f_a}

\title{Freedom and Constraints in the K3 Landscape}

\author{Vijay Kumar and Washington Taylor}

\affiliation{Center for Theoretical Physics\\
Department of Physics\\
Massachusetts Institute of Technology\\
77 Massachusetts Avenue\\
Cambridge, MA 02139, USA}

\emailAdd{vijayk@mit.edu}
\emailAdd{wati@mit.edu}

\preprint{MIT-CTP-4021}

\abstract{We consider ``magnetized brane'' compactifications of
the type I/heterotic string on K3 with $U(1)$ background fluxes.  The
nonabelian gauge group and matter content of the resulting six-dimensional vacua
are parameterized by a matrix encoding a lattice contained within
the even, self-dual lattice $\lat$.  Mathematical results of Nikulin
on lattice embeddings make possible a simple classification of  such
solutions.    This approach makes it possible to explicitly
and efficiently construct models in this class with a particular
allowed gauge group and matter content, so that one can immediately
``dial-a-model'' with desired properties.}

\begin{document}

\maketitle

\section{Introduction}

In this paper we consider a very simple class of string vacuum
solutions associated with supersymmetric 6-dimensional low-energy
physics.  Models of the type we consider here are characterized by an
intersection matrix associated with an even, self-dual lattice.  We use
theorems on lattice embeddings to give a characterization of
the space of allowed models.  This gives a concrete realization of a
piece of the landscape in which models can be conveniently
characterized and identified based on simple features of their
low-energy physics.  A brief summary and introduction to the main
technical results of the paper is given in Section \ref{sec:intro}.

While quite far from the physically observed world in which
supersymmetry is broken and space-time is 4-dimensional, the simple
mathematical structure described here may help provide insight into some
aspects of the larger space of string vacuum solutions, and in
particular into the question of whether low-energy physics is
constrained by string theory.  To put this work in context, a brief
discussion of the string vacuum problem and some of the motivation for
this work is given in Section \ref{sec:vacuum}.  Those uninterested in
landscape philosophy may skip Section \ref{sec:vacuum}.

\subsection{Introduction and outline of results}
 \label{sec:intro}

The models we consider in this paper are type I/heterotic string
theory compactifications on a K3 surface.  Such compactifications are
characterized by an $SO(32)$ bundle\footnote[3]{The case of the
$E_8\times E_8$ heterotic string can be treated in a completely
analogous manner.  The general story remains unchanged, but the
specific low-energy theories obtained are different.} over K3.  In
type I language, the background contains an orientifold 9-plane and 16
D9-branes whose charges combine to cancel the ten-form RR tadpole.  In
a curved background, such as a K3 surface, there is also a six-form
tadpole (D5-brane charge) \cite{bsv} that can be cancelled by
introducing world-volume fluxes for the gauge fields on the D9-branes.
These fluxes on the D9-branes thread 2-cycles in the K3 surface and
produce D5-brane charge equal to the instanton number of the flux
configuration \cite{smallinstantons, Douglas-branes}.

Models of the type considered here, with fluxes on D9-branes in type I
string theory encoding lower-dimensional D-branes wrapped on various
cycles, are known as ``magnetized brane models''.  Choosing different
fluxes on different D9-branes in these models leads to chiral fermions
in the dimensionally reduced theory \cite{gswest,
Bachas}.  These models were developed in the type I context in
\cite{magnetized} and have been widely studied in the literature.
Reviews of this work and further references appear in \cite{reviews}.

In general, the flux in the gauge fields on the D9-branes can be
nonabelian, but we restrict attention here to the simplified class of
configurations where only $U(1)$ fluxes, corresponding to fluxes on
individual D9-branes, are turned on.  Mathematically, this means that
the gauge bundle has a curvature that can be described in terms of a
direct sum of abelian connections on the individual D9-branes.  In the
more general case of nonabelian vector bundles a similar story holds,
but the technical details are more complicated.  
Recent related work on heterotic line bundle models appears in
\cite{heterotic-line, honecker, honeckertrap}.

On a torus, magnetized brane models are T-dual to intersecting brane
models \cite{IBM}; the D9-brane world-volume fluxes are T-dual to the
tilts of the diagonal intersecting branes on the dual torus.
(Intersecting brane models are also reviewed in \cite{reviews}.)  This
duality with intersecting brane models provides useful intuition for
the physics of magnetized brane models, although on a smooth K3, 
T-duality is replaced with mirror symmetry, which is harder to
describe explicitly.  Part of the motivation for this work was to
extend the systematic analysis  of
intersecting brane model solutions on toroidal orbifolds
to more general smooth Calabi-Yau spaces where intersecting brane
models are not particularly well defined, as discussed in Section 5
of \cite{Douglas-Taylor}.

For the magnetized brane models we consider here, each D9-brane
carries a separate $U(1)$ gauge field strength $F^i$.  This field
strength takes quantized values when integrated over any 2-cycle of K3
and thus, when suitably normalized by the unit of charge $\lambda$, lives in the integral 2-cohomology
of the K3 surface, {\it i.e.}, $\hat{f}_i= \lambda F^i/(2 \pi) \in H^2 (K3,\Z)$.
The cohomology group $H^2(K3,\Z)$ has the structure of a lattice, with
a natural inner product given by the wedge product in cohomology,
which is Poincar\'e dual to the intersection number of 2-cycles in
homology.  By a {\it lattice} in this paper we will always refer to a
vector space over $\Z$ carrying an integral, symmetric inner product.
For a K3 surface, the lattice defined by the cohomology group is the
unique even, self-dual unimodular lattice of signature (3,19), denoted
$\Gamma^{3, 19}$.  A review of the structure of this lattice and some
basic definitions relevant for lattices are given in \ref{app_lat}.

Any vacuum solution of this type is therefore associated with a set of
vectors $\hat{f}_i$ (not necessarily distinct) in the lattice
$\Gamma^{3, 19}$.  Since we have turned on $U(1)$ fluxes, the gauge
group in the 6D theory is broken from $SO(32)$ to the commutant of the
$U(1)$ fluxes.  If we were to turn on 16 distinct $\hat{f}_i$'s the
gauge group would be broken to $U(1)^{16}$.  The tadpole constraint in
fact restricts the number of nonzero $\hat{f}_i$'s to be at most 12 for smooth K3 geometries.
Note that the vectors $\hat{f}_1,\cdots, \hat{f}_{16}$ need not be
distinct, and some of them could be equal.  When some of the
$\hat{f}_i's$ are equal, the gauge group on the branes is expanded to
a nonabelian group.  Let $f_a$ denote the distinct, non-zero
vectors among the set
$\{\hat{f}_1,\hat{f}_2,\cdots,\hat{f}_{16}\}$. We use the $\hat{f}_i$
(with hats) to denote the fluxes on each D9 brane, where the index $i$
runs over individual branes, and the notation $f_a$ (no hats) to
denote the distinct fluxes among these. In the most general situation,
we have $K$ {\it magnetized stacks} of $N_a$ equal fluxes $f_a,\ 
a=1,\cdots,K$ where $f_a\neq 0$ and all the $f_a$ are distinct.  By a
{\it magnetized stack} we mean a set of $N_a$ D9-branes each with the
same value of (non-zero) flux $f_a$ turned on. The stack of branes
with $\hat{f}_i=0$ will be called the {\it unmagnetized stack}.

The gauge group and matter content of the dimensionally reduced 6D
theory is completely determined by the sizes $N_a$ of the flux stacks
and the topological invariants $f_a\cdot f_b$ of the fluxes, where
$\cdot$ denotes the inner product on the lattice. The gauge group and
matter content can be enhanced when the K3 surface becomes singular.
We focus in this paper on smooth K3 compactifications, where
the gauge group is given by a product of unitary groups
$\prod_a U(N_a)\times SO(32-2\sum_a N_a)$; the unitary groups
arise from the magnetized stacks and the orthogonal group arises from
the unmagnetized stack.  (Note that generally the $U(1)$ factors in
$U(N_a)$ are anomalous and get masses through the Green-Schwarz/Stueckelberg
mechanism).  The number of matter fields in the various allowed
representations of the gauge group can be written in terms of the {\it
intersection matrix} $m_{ab}:=f_a\cdot f_b$ and the stack sizes
$\{N_a\}$.  Thus, the stack sizes and intersection matrix parameterize
the possible low-energy theories realizable in this construction.

To understand the range of models possible, we therefore need to know
what intersection matrices $m_{ab}$ and multiplicities $N_a$ can be
realized in string theory.  This question can be conveniently
reformulated in the language of lattice embeddings.  We can think of
the vectors $\{f_{a=1,\cdots ,K}, \frac{1}{2}\sum_a N_a f_a\}$ as generating 
a lattice $\M$ of rank $\ \leq \ K$  with inner product determined
by the matrix $m$.\footnote[4]{The lattice $\M$ is required
to contain the integral vector $\fwvec$ to avoid global anomalies. For a
smooth K3 compactification, we also require $\M$ to not have any vectors
with norm-squared $-2$.}  
The main result of this paper is the
conclusion that existing theorems on lattice embeddings, due to
Nikulin and earlier authors, provide a fairly complete characterization of
the space of abelian magnetized brane models on K3 through the
classification of allowed intersection matrices $m$.  The vanishing of
the six-form R-R tadpole gives a constraint $\sum_i \hat{f}_i\cdot
\hat{f}_i = {\rm Tr}\; \hat{m} = -48$ on the trace of the full
intersection matrix $\hat{m}_{ij} = \hat{f}_i \cdot \hat{f}_j$.  This
becomes $\sum_{a} N_a f_a \cdot f_a = -48$ when we rewrite the sum
over branes into one over stacks.  Low-energy supersymmetry further
requires the lattice $\M$ to be negative-definite.  Up to these
constraints, Nikulin's results on lattice embeddings essentially state
that {\it any} desired intersection matrix can be realized in a K3
compactification, except possibly in cases where the rank of the matrix is 11 or 12.
Furthermore, the realization of a specific
intersection matrix is almost unique up to automorphism (relabeling)
of the K3 cohomology lattice, with the ``almost'' caveat arising from
some discrete redundancies\footnote[5]{Equivalent
discrete redundancies were recently encountered in analysis of dyon
models \cite{Sen-dyon}.}
 associated with a finite number of possible
refinements (overlattices) of the lattice $\M$.

The upshot of this analysis is that for this family of models the
intersection matrix $m$ provides a complete means of
classifying the range of possible models according to simple
features of their low-energy physics.  The methods described here can be used to enumerate
the models in the class in a straightforward way, either to search for
models with specific features or to perform a statistical analysis on
the class of models.  Furthermore, the intersection matrix $m$
provides a straightforward way to directly construct all models with a
particular desired feature.  For example, one might wish to construct
all models with a particular group, like $G = SU(3) \times SU(2)$
as a subgroup of the full group and at least 3 chiral
fermions in the bifundamental $(\bf{3},\bf{2})$.  This can be
formulated as a constraint on the intersection matrix $m$, so that all
models with the desired feature can immediately be constructed or
enumerated.  Again, this could be useful either for model-building
purposes or for statistical analysis of the space of models.

The analysis of this paper thus gives a particularly simple example of
a region of the landscape where models can be easily classified and
explicitly constructed.  It is interesting to contrast this with other
types of models where classifying all allowed vacuum solutions is
difficult or impossible with current methods.  In some cases,
identifying the complete set of models with a particular set of
physical features may be computationally difficult or intractable.
For example, consider the closely related family of intersecting brane
models on a toroidal orientifold.  The simple and popular toroidal
model $T^6/\Z_2 \times \Z_2$ \cite{toroidal-orbifold} has been studied
in many papers over the last decade and gives rise to semi-realistic
4D theories with a gauge group containing the standard model gauge
group and 3 generations of chiral matter fields \cite{csu, cim, cl3, Marchesano-Shiu, cll, reviews}.  Due,
however, to the presence of supersymmetric branes with tadpole
contributions of mixed signs in this class of models, a complete
understanding of the space of allowed models poses a significant
challenge.  In \cite{bghlw}, a systematic search using a year of
computer time sampled a large number of vacua in this class.  In
\cite{Douglas-Taylor}, it was proven that the set of inequivalent
models in this class is finite.  The classification of these models
has now been completed and will appear in \cite{Rosenhaus-Taylor}.  In
this toroidal orbifold construction, finding the set of models with a
particular physical feature can be quite a nontrivial problem.  (For
example, finding all models with gauge group containing the subgroup
$SU(3) \times SU(2)$ poses a significant computational challenge,
whose solution appears in \cite{Rosenhaus-Taylor}).  In the class of
K3 models considered in the present paper, on the other hand, such a
construction is relatively straightforward.  The constraint on the gauge group
can be imposed on the stack sizes $N_a$ and intersection matrix
$m_{ab}=f_a \cdot f_b$, and all $N_a,m_{ab}$ satisfying this
constraint can immediately be constructed and tabulated (as discussed in
Section \ref{sec:dialamodel}).

In the following subsection we summarize some outstanding issues
related to the large landscape of string vacua and the problem of
predictability in this context.  In brief, the message of this paper
is that in this special corner of the string landscape the space of
string theory compactifications has a very simple organizing
principle.  
Given the parameterization of the allowed space of models
in terms of the intersection matrix $m$, 
in this region of the landscape there is both freedom and constraint, 
in that \emph{any} negative-semidefinite matrix $m$ that determines
an even lattice $\M$ of rank $\leq 10$
can be realized in string
theory, provided that $\M$ has no $-2$ vectors and satisfies
the linear tadpole constraint.  Furthermore, up to some limited discrete redundancy, any
allowed low-energy physics structure is realized in a unique way up to
dualities.  While such a simple framework is unlikely to extend to the
whole string landscape in any foreseeable fashion, this gives an
example of a part of the landscape where underlying mathematical
structure gives great simplicity to the discrete variety of allowed
string theory solutions.

The structure of the rest of this paper is as follows: In Section
\ref{sec:basics} we review the basic geometry of K3.  In Section
\ref{sec:models} we describe the class of type I magnetized brane
models of interest.  The 6D dimensionally reduced physics of these
models is described in Section \ref{sec:6D}.  The existence of models
for any negative-semidefinite even matrix of rank $\leq 10$ that satisfies
the tadpole constraint is demonstrated using Nikulin's theorems in
Section \ref{sec:theory}.  A systematic approach to enumeration of
models including the determination of discrete redundancies is
presented in Section \ref{sec:enumeration}, along with some explicit
examples.  In Section \ref{sec:dialamodel} we show how all models with
particular desirable features such as a specific low-energy gauge
group or matter content can be efficiently enumerated.  Concluding
remarks appear in Section \ref{sec:conclusions}.  Appendices contain
additional technical material on the structure of the lattice
$\lattice$ and Nikulin's lattice embedding theorems.

\vspace{0.1in}
\noindent
{\bf Note for version v4}: We have made the following significant
changes to this version: a factor of two error in the tadpole
condition has been corrected, we have restricted ourselves to smooth
K3 compactifications by eliminating vacua that involve $-2$ vectors in
the Picard lattice of the K3 surface, and the mod 2 constraint from
the Freed-Witten global anomaly has been included. These modifications
do not affect the central conclusion of this paper that string vacua
in this part of the landscape can be characterized through the existence of lattice
embeddings. However, some  specific quantitative statements
regarding enumeration of models
have been revised. We are grateful to Edward Witten, Ilarion Melnikov
and the authors of \cite{Ilarion} for helpful communications regarding these issues.

\subsection{Predictions and the vacuum problem}
\label{sec:vacuum}

It has been known since the early days of string theory that the space
of supersymmetric string vacua is vast.  For example, for any
Calabi-Yau complex 3-manifold there is a continuous moduli space of
supersymmetric string vacuum solutions.  More recently, the
experimental observation of a positive nonzero cosmological constant
has led to a new effort to understand the ``landscape'' of metastable
string vacua with positive cosmological constant.  Incorporation of
fluxes and other mechanisms can stabilize the continuous moduli of a
space of vacuum solutions, and in principle with the breaking of
supersymmetry the cosmological constant can be lifted to a positive
value.  The exponentially large number of ways in which fluxes can be
introduced gives an exponentially large number of candidate string
vacua, whose enormous multiplicity has been invoked as an explanation
for the existence of solutions having a positive cosmological constant
with the observed small value.  Reviews of developments on the
landscape of flux compactifications appear in \cite{reviews-fluxes}.

In considering the apparent enormous multiplicity of stable and
metastable string theory vacuum solutions, the problem of extracting
predictions for the low-energy theory becomes increasingly acute.  It
has been suggested that statistical analysis of the space of allowed
models may provide fruitful insights \cite{statistics}.  In the
absence of a true understanding of background-independent dynamics of
string theory, however, it is as yet very difficult to extract physically
meaningful predictions from the statistics of vacua or to make
progress on other proposed vacuum selection mechanisms.

Thus, without some dramatic new insight into the global and temporal
dynamics of string theory, at the present the clearest course of
action is to consider the full space of possible string solutions.  It
is possible that string theory imposes absolute constraints on
low-energy physics.  If present, such constraints should be observed
in any sufficiently broad family of string vacua.  Here we use a very
simple framework to illustrate the question (more statistical
approaches to this kind of question have recently been taken in
\cite{bbn, Dienes}; for a more bottom-up perspective see \cite{Kane}).
Given some set of string vacua ${\cal V}$ (for example, intersecting
brane models on a particular toroidal orbifold), and a set of
low-energy observables ${\cal O}$ (for example, the gauge group,
matter content, Yukawa couplings etc.), there are roughly two extremes
according to which the observables can be distributed in the set of
vacua.

\noindent
{\bf A)} ``Anything goes''.  In this extreme, essentially all possible
values of the observables ${\cal O}$ in question can be realized
within the set of vacua ${\cal V}$.  For continuous observables like
masses and couplings, this might mean that any possible combination of
masses and couplings could be realized to a particular level of
precision dependent on the number of vacua in the set ${\cal V}$.  (In
some cases there may be a fixed range within which this freedom is
realized, so that distributions of type {\bf A} may really be best
described as {\bf B} ``bounded'')

\noindent
{\bf C)} ``Constraints''.  In this extreme, there are absolute
constraints on the set of observables.  For example, one could imagine
that for a given set of vacua, any model that contains $SU(3) \times
SU(2)$ has precisely 3 generations of bifundamental ``quark-like''
matter fields.  Or one might find that a constraint on a combination
of several masses and couplings gives one of these parameters as a
function of the others.

To make a specific quantitative prediction from string theory one
would need to show that for all possible sets of vacua, a common
constraint of type {\bf C} arises in the low-energy theory.  Of
course, the set of vacua within our current range of understanding is
probably only a tiny fraction of the full range of string solutions,
so we cannot reasonably show this for all sets of vacua.  If, however,
we can demonstrate that in a number of ostensibly unrelated corners of
the landscape there are common constraints on the low-energy physics,
this would represent a hypothetical universal constraint, which could
be tested by examining other classes of string solutions.

On the other hand, to show that a given set of observables, taken in
isolation, is not constrained by string theory, it suffices to find a
single class of string models in which situation {\bf A} holds.  For
example, in \cite{bghlw, Douglas-Taylor, Rosenhaus-Taylor} the space
of supersymmetric intersecting brane models on the $T^6/\Z_2
\times\Z_2$ toroidal orientifold was studied systematically.
Considering only the low-energy gauge group and numbers of matter
fields in different representations, this class of models seems to
give a distribution best described by {\bf A)}, ``anything goes''.
Although there is a maximum to the size of the gauge group and number
of matter fields that bounds the range of the distribution and
produces a sparse ``tail'', small groups like $U(N) \times U(M)$, $N,
M \leq 5$ can all be realized in many ways as part of the full gauge
group in this construction with arbitrary (but not too large) numbers
of chiral matter fields in the $(N, \bar{M})$ representation.  This
confirms, as has long been known, that string theory cannot uniquely
determine the gauge group and number of generations of matter without
additional constraints (like perhaps SUSY breaking).  Of course, for
this particular orientifold construction, the total number of possible
models is somewhat limited.  As shown in \cite{Rosenhaus-Taylor}, the
number of distinct ways the gauge group $SU(3) \times SU(2)$ can be
realized as a subgroup of the gauge group with 3 generations of
``quarks'' in this class of models is limited to several thousand.
Thus, the ``anything goes'' nature of the general distribution is
limited by the finite sample size.  Indeed, while a number of
semi-realistic models have been found in this framework \cite{csu,
cim, cl3, Marchesano-Shiu, cll}, these generally have extra massless
chiral exotic fields, and differ from the standard model in observable
ways.  A larger sample is needed to reproduce more detailed features
of the standard model, even in the absence of definitive stringy
constraints.  Such larger samples can be found, for example, by
generalizing to arbitrary Calabi-Yau geometries or Gepner model
constructions \cite{Schellekens}.

To date there are no conclusively demonstrated broad constraints on
the space of low-energy theories that can be realized in string
models (though some interesting suggestions leading towards such
constraints have been made in \cite{swampland}).  Thus, in order to
see whether string theory produces constraints on low-energy theory
{\it anywhere} in the landscape it may be fruitful to begin with the
simplest classes of string compactifications.  The string vacuum
solutions we consider here, associated with type I magnetized branes
on K3, live in such a simple, and hopefully easier to analyze, corner
of theory space.  The resulting vacua, which are 6-dimensional and
supersymmetric, are far from the physically observed world.  
These models obey some specific constraints, but can vary freely subject to these constraints
Specifically, we find that the gauge group and matter content
of the low-energy theory have a particular dependence on the stack
sizes and intersection matrix parameters $N_a, m_{ab}$.  
\emph{Any} intersection matrix $m_{ab}\ $ subject to the constraint that the lattice $\M$ it determines is even, negative-definite, has no norm-squared $-2$ vectors, and satisfies a simple tadpole constraint, can be realized in an almost unique way as a string compactification, at least in cases when rank $m \neq 11, 12$.

The simple mathematical structure found for this class of models, and
its consequences for the part of the landscape where this analysis is
relevant, suggests that a search for a similar decomposition of the
landscape distribution into constraints and freedom may be helpful in
a broader class of string vacua.  Such analysis might help to address
the question of whether, even in a simple supersymmetric and
higher-dimensional context, string theory has the potential to
constrain any part of the low-energy field theory arising from
compactification.

\section{Some basic facts about K3 surfaces} 
\label{sec:basics}

In this section we review the basic properties of K3 surfaces that are
relevant to our discussion.  A more detailed exposition complete with
references can be found in Aspinwall's lectures on K3 surfaces
\cite{Asp96a}.  A K3 surface $S$ can be defined as a compact, two
complex dimensional, simply-connected Calabi-Yau manifold.  In other
words, $S$ is a compact, simply-connected complex surface with trivial
canonical bundle.  Like all Calabi-Yau manifolds, any K3 surface
admits a Ricci-flat K\"ahler metric.  The reason Calabi-Yau manifolds
are of interest is that they are an exact solution of the superstring
equations of motion and they preserve some of the supersymmetry of the
ten-dimensional theory.  There are many thousands of topologically
distinct Calabi-Yau three-folds known \cite{ks} and it is not yet
clear whether the total number of such manifolds is finite or
infinite.  Two-dimensional compact Calabi-Yau manifolds are severely
constrained in this regard and in fact there is only one
simply-connected Calabi-Yau surface (the K3 surface) up to
diffeomorphism.  Thus, all K3 surfaces have the same topology which
can be determined by studying any specific one.  In particular, the
Betti numbers $b_n(S)=\mbox{dim}\ H_n(S,\Z)$ are completely determined
and they are $b_1=b_3=0, b_2=22$.  For a K3 surface $\HZ$ is
torsion-free and therefore has the structure of a lattice, by which we
mean a free $\Z$-module, i.e $\HZ \cong \Z^{22}$ (as a free abelian group).

The lattice $\HZ$ has a natural integral, symmetric, bilinear form
$\Phi(x,y)$ defined as the intersection number of the two 2-cycles
$x,y\in \HZ$.  The intersection number is a well-defined quantity on
homology classes \cite{griffithsharris} and satisfies $\Phi(x, y) \in
\Z$ (integral), $\Phi(x, y) = \Phi(y, x)$ (symmetric) and $\Phi(x, m
y+nz)=\Phi(my+nz, x) = m\Phi(x, y)+n\Phi(x, z)$ (bilinear) for all
$x,y,z \in \HZ$ and $m,n \in \Z$.  A lattice $\L$ with an integral,
symmetric, bilinear form is said to be {\it even} if for all $x,y\in
\L$, $\Phi(x,y)$ is an even integer.  $\L$ is {\it self-dual} if $\L
\cong \L^*:= \mbox{Hom}(\L,\Z)$.  Given a basis $\{e_\alpha\}$ for the
lattice $\L$, the bilinear form defines a matrix
$C_{\alpha\beta}:=\Phi(e_\alpha, e_\beta)$.  The {\it signature} of
$\L$ is defined as $(l_+,l_-)$, where $l_+$ and $l_-$ are the number
of positive and negative eigenvalues of $C$ respectively.  

It can be shown that the lattice $\HZ$ (with the bilinear form $\Phi$)
associated with any K3 surface $S$ is an {\it even}, {\it self-dual}
lattice of {\it signature} $(3,19)$ \cite{Asp96a}.  In fact, there is
precisely one such lattice with these properties up to
isometry\footnote[3]{Two lattices $\L$ and $\L'$ are {\it isomorphic}
if there exists a bijective map $f:\L\rightarrow \L'$.  If $f$
preserves the bilinear form $\Phi$ then $f$ is said to be an {\it
isometry}.}; this lattice is denoted by $\lat$ (See Appendix \ref{app_lat}).  Poincar\'e duality
implies that the lattices $H^2(S,\Z)$ and $\HZ$ are
isometric\footnote[5]{The Poincar\'e dual of a 2-cycle $A\in \HZ$ is
a closed two-form $\tilde{A} \in H^2(S,\Z)$ such that for every
2-cycle C, $\int_C \tilde{A} = \Phi(A,C)$.  Poincar\'e duality maps
the intersection product $\Phi$ in homology to the wedge product of
forms in cohomology \cite{griffithsharris}.  Given closed 2-cycles
$A,B$ and their Poincar\'e duals $\tilde{A},\tilde{B}$, we have
$\int_S\tilde{A}\wedge \tilde{B} = \Phi(A,B)$.} and that $H^2(S,\Z)
\cong \lat$.  For the rest of the paper, we will use the notation
$x\cdot y$ to denote the bilinear form on a lattice, which in this
case would be the intersection number in homology, or the wedge
product of forms in cohomology.

Given a complex structure on $S$, there exists a globally defined,
nowhere vanishing, holomorphic two form $\Omega$.  A complex structure
defines a Hodge decomposition of the lattice $H^2(S,\Z) \subset
H^2 (S,\C)$ as
$H^2(S,\C) = H^{2,0}(S)\oplus H^{1,1}(S) \oplus H^{0,2}(S)$.
$\Omega \in H^{2,0}(S)$ and is defined only up to scaling by a
non-zero complex number.   The Hodge decomposition determined by
$\Omega$ uniquely specifies the complex structure.  We can write
$\Omega=x+iy$, where $x,y \in H^2(S,\R)$.
The
elements $x$ and $y$ are constrained because $\Omega$ satisfies 
\begin{equation}
\int \Omega \wedge \Omega = 0, \quad \int \Omega \wedge \bar{\Omega}
\propto \mbox{Vol}(S) > 0  
\end{equation}
Therefore, $x$ and $y$ satisfy $x\cdot x = y \cdot y > 0$ and $x \cdot
y = 0$.  These conditions imply that the real plane spanned by $x$ and
$y$ in $H^2(S,\R) \cong \R^{3,19}$ is positive definite.  Thus, the
complex structure on $S$ is determined by a positive-definite two-plane in the
space $\R^{3,19}  \supset H^2(S,\Z)$.
The K\"ahler form $J$ is a real, closed $(1,1)$ form and satisfies
\begin{eqnarray}
\int J\wedge \Omega & = & 0 \Rightarrow J\cdot x = J \cdot y =
0\nonumber \\
\int J \wedge J & \propto & \mbox{Vol}(S) > 0 \Rightarrow J\cdot J > 0
\end{eqnarray}
The above equations imply that the holomorphic two-form and the K\"ahler form determine a positive-definite three-plane in $\R^{3,19}$.  

\section{Type I/Heterotic on K3}

\label{sec:models}

We wish to construct $d=6$ vacua with chiral $(1,0)$ supersymmetry (8
real supercharges) starting with the type I or heterotic string theory
on a K3 surface $S$.  The type I and $SO(32)$ heterotic
string\footnote[3]{The gauge group is really Spin(32)/$\Z_2$ and not
$SO(32)$ for both the type I and the heterotic string.  See
\cite{Berkoozetal, orientifold-vector} for further discussions of this
point.} theories are equivalent under a
strong-weak duality \cite{heteroticduality}.  The low-energy effective
action of both string theories is type I supergravity.  Since our
analysis is carried out using the effective action alone, it applies
to both the type I and the heterotic string.  Some calculations are more
transparent in the heterotic string, while the type I string gives an
intuitive picture of the $SO(32)$ gauge group in terms of D9-branes.  In
this section we discuss the constraints imposed by tadpole
cancellation and low-energy supersymmetry.  Readers familiar with
these constraints can skip to the summary at the end of this section.

\subsection{Tadpole cancellation}

In the context of the heterotic string, the tadpole constraint arises
from integrating the Bianchi identity for the NS-NS $B$ field over the
K3 surface.  In the type I theory the same constraint arises from the
equation of motion for the six-form RR potential, including the
Wess-Zumino couplings of the O9-plane and the D9-brane in the presence
of world-volume fluxes.  Since these are dual descriptions of the same
system, both approaches must yield the same answer.  We summarize the
calculation of the constraint here in heterotic language, which is
perhaps more transparent, following \cite{strominger}.  We then
summarize the physical content of this constraint as interpreted in
the type I theory.

The Bianchi identity in the heterotic string is \cite{gsw2}
\begin{equation}
dH = \mathrm{tr}\, R\wedge R - \mathrm{tr}\, \CF \wedge \CF\, ,
\end{equation}
where $R$ and $\CF$ denote the curvatures of the tangent bundle and gauge bundle respectively, and the trace is to be taken in the vector representation. Integrating this equation over $S$ we obtain 
\begin{equation}
-\frac{1}{8\pi^2}\int_S \mathrm{tr}\, \CF \wedge \CF = -\frac{1}{8\pi^2}\int_S \mathrm{tr}\, R\wedge R \, .
\end{equation}
This is an equality of the Pontyagin classes of the gauge bundle and tangent bundle integrated over $S$. For a K3 surface the right hand side evaluates to -48 \cite{eguchi}, and this leaves us with the tadpole constraint
\begin{equation}
\frac{1}{8\pi^2}\int_S \mathrm{tr}\, \CF \wedge \CF = 48 \, .
\label{eq:tadpole1}
\end{equation}

In the type I language, the orientifold action fixing all
space-time points gives an O9-plane with ten-form R-R charge $-32$.  The
$SO(32)$ gauge group lives on the 16 D9-branes which, along with their
orientifold images, must be added to cancel the R-R charge from the
O9-plane.  Due to the non-zero Riemann curvature of S, the
gravitational couplings of the  D9-branes and
O9-plane \cite{bsv,wzcouplings} induce a net nonvanishing
six-form R-R tadpole.  We cancel the tadpole by turning on background
world-volume fluxes with the appropriate instanton number on the
D9-branes.  These world-volume fluxes on the D9-branes thread
2-cycles in S and have D5-brane charge equal to the instanton number
of the flux configuration \cite{smallinstantons, Douglas-branes}.  

\subsection{$U(1)$ bundles and Dirac quantization}

We restrict our attention to gauge bundles that are sums of $U(1)$
bundles for the sake of simplicity.  $SO(32)$ is a rank 16 group and
has sixteen $U(1)$ factors.  Let $T^i, \ i=1,2,\cdots ,16$ denote the
sixteen anti-Hermitian Cartan generators in $so(32)$ normalized so that 
$\mathrm{tr}\, (T^i T^j) = -2\lambda^2 \delta^{ij}$. The curvature form 
can be expanded in this basis as $\CF = F^i\, T^i$ where 
$F^i\in H^2(S,\R)$ are closed 2-forms, some of which could be zero.  
We know that $U(1)$ fluxes threading compact 2-cycles inside $S$ are 
subject to Dirac quantization. With our normalization conventions this 
implies the integrality of the class 
\begin{equation}
\hat{f}_i := \frac{\lambda }{2\pi} F^i \in H^2(S,\Z)\, .
\end{equation}
In terms of the $\hat{f}_i$, the tadpole condition (\ref{eq:tadpole1}) 
reads
\begin{equation}
\sum_{i=1}^{16} \hat{f}_i\cdot \hat{f}_i = -48\, . \label{eq:tadpole}
\end{equation}
Here the $\cdot$ denotes the intersection product on the lattice
$H^2(S,\Z)$ as explained in Section \ref{sec:basics}.

It is useful to keep in mind the interpretation in terms of the type I
theory.  A single D9-brane with a $U(1)$ world-volume flux $F^i$ has
an orientifold image with flux $-F^i$.  The brane and its image
together form a single, dynamical type I D9-brane with $SO(2)$
Chan-Paton indices.  The type I theory contains sixteen D9-branes with
$SO(32)$ Chan Paton indices and we therefore can turn on a maximum of
sixteen independent $U(1)$ fluxes $F^i, \ i=1,2,\cdots, 16$.

In addition, to avoid global anomalies we must impose that the first Chern
class of the gauge bundle be an even integral class on $S$ 
\cite{Witten-global, Freed}.  Therefore, 
\begin{equation}
\frac{1}{2} \sum_i \hat{f}_i \in H^2(S,\Z)\, .
\end{equation}

\subsection{Supersymmetry constraint}
\label{sec:susycond}

Since we demand that the six-dimensional vacua have supersymmetry, the
$SO(32)$ gauge bundle is constrained.  It is well known \cite{gsw2, marino}
that to preserve supersymmetry the curvature form $\CF$ must satisy
$\CF_{\alpha\beta}=\CF_{\bar{\alpha}\bar{\beta}}=0, \
\CF_{\alpha\bar{\beta}}\ g^{\alpha\bar{\beta}} = 0$.  Here $\alpha$
and $\beta$ are holomorphic coordinates on $S$ and
$g^{\alpha\bar{\beta}}$ is the K\"ahler metric.  In terms of the
$U(1)$ fluxes, this implies
$F^i_{\alpha\beta}=F^i_{\bar{\alpha}\bar{\beta}}=0, \
F^i_{\alpha\bar{\beta}}\ g^{\alpha\bar{\beta}}=0$ for each $i$.  The
first two conditions are equivalent to the global conditions
\begin{equation}
\int F^i\wedge \Omega = \int F^i\wedge \bar{\Omega} = 0.  
\label{eq:global-conditions}
\end{equation}
This is because on a K3 surface, any (2,0) form is proportional to
$\Omega$ up to a complex constant; the integral conditions
(\ref{eq:global-conditions})
imply that
this constant of proportionality is zero, and therefore that the local
conditions are satisfied.  This implies that $F^i$ is a $(1,1)$-form, and therefore that the corresponding $U(1)$ bundle is holomorphic. Given a holomorphic $U(1)$ bundle with curvature form $F^i$,
the third (local) condition $F_{a\bar{b}}\ g^{a\bar{b}}=0$ is
satisfied iff the global constraint $\int F^i \wedge J = 0$ is true
\cite{Strominger:1985ks}.  Thus, supersymmetry places three (global) conditions on the
curvature two-form $F^i$
\begin{equation}
\int F^i\wedge \Omega = 0 \quad \int F^i\wedge \bar{\Omega} = 0 \quad
\int F^i\wedge J=0 \label{eq:SUSY-conditions}
\end{equation}

We now ask the question --- ``Given a set of line bundles on K3
specified by the lattice vectors $\{\hat{f}_i\}$ in $\lattice$, does
there exist a choice of $\Omega$ and $J$ so that supersymmetry is
preserved?''.  We know from our discussion in Section \ref{sec:basics}
that $\Omega$ and $J$ span a positive-definite three-plane in the
space $\R^{3,19}$ $= H^2(S,\R)$.  Since the lattice has signature
(3,19), if we restrict the $\{\hat{f}_i\}$ to a negative-definite
plane $\Sigma \subset\R^{3, 19}$ (dimension $\leq 16$), there always
exists a real, positive definite, 3-plane orthogonal to it.  We can
always choose $\Omega$ and $J$ to lie in the positive definite
3-plane. Therefore, the only
constraint from supersymmetry is that the lattice vectors
$\{\hat{f}_i\}, i=1,2,\cdots, 16$ generate a negative-definite
sublattice in $H^2(S,\Z)$.  This implies in particular that
$\hat{f}_i^2 < 0$.  Since the lattice is even, we have $\hat{f}_i^2
\leq -2$.  
When $\hat{f}_i^2 = -2$ the Poincar\'e dual homology class $[C_i] \in H_2(S,\Z)$, or its negative $-[C_i]$, contain an irreducible rational curve \cite{Asp96a}. The supersymmetry conditions \eqref{eq:SUSY-conditions} require that this holomorphic curve have zero volume, thereby rendering the K3 surface $S$ singular. In order that $S$ be smooth we must require that the lattice $H^2(S,\Z) \cap H^{1,1}(S)$, known as the Picard lattice, not contain any norm-squared $-2$ vectors. This can be ensured by demanding that the real vector space spanned by the $\hat{f}_i$ (a sub-space of $\mathbb{R}^{3,19}$) not contain any integral vectors with norm-squared $-2$. In particular, this implies that $\hat{f}_i^2 \leq -4$.
Note that the vectors $\hat{f}_i$ need not be linearly
independent, or even distinct, and so are not necessarily a basis of
minimal dimension for the lattice they generate.  The tadpole
constraint (\ref{eq:tadpole}), $\sum_i\hat{f}_i^2 = -48$, can only be
satisfied if at least four $\hat{f}_i$'s are zero.  Therefore, even
though the gauge group $SO(32)$ has sixteen $U(1)$ factors, at most
twelve of them can be turned on.

\subsection{Summary of constraints}
The tadpole condition (\ref{eq:tadpole1}) can be satisfied by turning
on background world-volume fluxes threading various 2-cycles in the
K3.  We focus on $U(1)$ world-volume fluxes for simplicity.  Dirac
quantization implies that the $U(1)$ fluxes, appropriately normalized,
are vectors in the integral cohomology lattice $H^2(S,\Z)\cong \lat$.
For each $U(1)$ factor, we have a lattice vector $\hat{f}_i$; in terms
of these vectors $\hat{f}_i$, the tadpole condition is simply $\sum_i
\hat{f}_i\cdot \hat{f}_i=-48$.  
Supersymmetry requires that the non-zero vectors $\hat{f}_i$ (not
required to be linearly independent, or distinct) generate a
negative-definite sublattice of $\lat$.  The K3 surface is smooth only if the
lattice $H^2(S,\Z) \cap H^{1,1}(S)$ contains no norm-squared $-2$ vectors, which implies
that $\hat{f}_i^2 \leq -4$. This in turn implies that we can have at most 
twelve non-zero $\hat{f}_i$. Let $\M$ denote the lattice generated (over $\Z$)
by the vectors $\hat{f}_i$ and $(1/2)\sum_i \hat{f}_i$. Then,
\begin{align}
1)  &\ \M  \mbox{ is a rank $\leq 12$ negative-definite sublattice
  of } \lattice\, , \nonumber \\
2)  & \ \mbox{The vector space } \R \otimes \M \mbox{ contains no integral norm-squared $-2$ vectors in $\lat$,} \label{conditions} \\ 
3)  & \ \sum_i \hat{f}_i\cdot \hat{f}_i = -48  \nonumber \, .
\end{align}
When these constraints are satisfied, we have a consistent 
string compactification on a smooth K3 surface that satisfies the equations of motion in the supergravity approximation. 
That is, for any set of fluxes
$\hat{f}_i \in\lattice$ satisfying (\ref{conditions}), there exist 
choices of complex structure $\Omega$ and K\"ahler form $J$ on K3 satisfying
(\ref{eq:SUSY-conditions}).  The moduli space of these
solutions is restricted by the choice of fluxes, since the complex
structure and K\"ahler form of the K3 surface are constrained to be
orthogonal to the lattice generated by the fluxes $\hat{f}_i$. While we have restricted our attention to smooth K3 compactifications, it is possible that some of the singular compactifications are consistent; we comment on this in the following section.

\section{Six-Dimensional Physics}
\label{sec:6D}

In this section we describe how the parameters of the compactification
discussed in the previous section affect the low-energy physics.
These six-dimensional, supersymmetric compactifications of the $SO(32)$
string on K3 have been studied earlier in \cite{gswest,
honecker, honeckertrap}.  In six-dimensional theories
with $(1,0)$ supersymmetry, there are four kinds of massless
multiplets (see Appendix B of \cite{polchi2}), listed in Table \ref{table1}.
\begin{table}
\centering
\begin{tabular}{|c|c|}
\hline 
Multiplet & Matter Content \\
\hline
SUGRA & $(g_{\mu \nu}, B^+_{\mu \nu}, \psi^-_{\mu})$ \\
Tensor & $(B^-_{\mu \nu}, \phi, \chi^+)$\\
Vector & $(A_\mu, \lambda^-)$ \\
Hyper & $(4\phi, \psi^+)$ \\
\hline
\end{tabular}
\caption{SUSY representations in d=6 with 8 supercharges.  The + and -
indicate the chirality for fermions and self-duality or
anti-self-duality for the two index tensor.} \label{table1}
\end{table}
The SUGRA and the tensor multiplets are uncharged under the gauge group, and the vector multiplet transforms in the adjoint.  In six dimensions, the CPT conjugate spinor has the same chirality as the spinor.  This implies that chiral fermions must always be charged under a real representation of the gauge group.  The (chiral) hypermultiplet therefore transforms under a real representation.  

\subsection{Gauge group and matter content} \label{matter}
Turning on $U(1)$ fluxes breaks the $SO(32)$ down to the commutant of
the $T^i$ for which $F^i\neq 0$.  As discussed in Section
\ref{sec:susycond}, supersymmetry allows us to turn on at most twelve
$F^i$.  Since at least four of the $F^i$ are zero, the gauge group
always has an unbroken $SO(8)$ subgroup.  If we choose twelve distinct
nonzero $F^i$'s, the gauge group is broken to $U(1)^{12}\times SO(8)$.
When some of the $F^i$'s are equal to each other one obtains unitary
groups.  In the type I picture, this would mean that a stack of
several D9-branes have the same background $U(1)$ flux turned on.  For
example, with $F_1=F_2=\cdots=F_7\neq 0$ and $F_8=F_9=\cdots
=F_{16}=0$, the commutant is $U(7)\times SO(18)$.  In general, the
gauge group in these compactifications is $\prod_{a=1}^{K} U(N_a)
\times SO(2M)$ with $N_1+N_2+\cdots+N_K+M=16$.  This pattern of
breaking is accomplished by having $K$ stacks, each with $N_a$
D9-branes where $a$ is an index for the stacks $a=1,\cdots,K$.  The
$U(1)$ factors in the gauge group are generally anomalous due to the
couplings with the two-form potential $B_{\mu\nu}$. However, in some
cases, there are linear combinations of the $U(1)$ factors that remain
anomaly-free. In what follows we do not address the question of
precisely which $U(1)$ are lifted, and we simply retain the $U(1)$
factors in the gauge group as it does not affect the classification
and enumeration of vacuum solutions.  We reserve the indices $i,j$ to
denote individual D9-branes and the indices $a,b$ to denote stacks of
branes.
\begin{figure}
\centering
\includegraphics[width=3in]{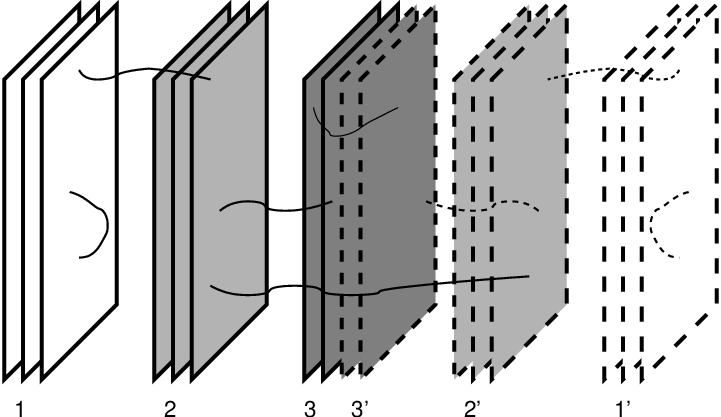}
\caption{Three stacks of D9-branes and their orientifold images.  The
$1$ and $2$ stacks with $N_1$ and $N_2$ branes respectively have
$U(1)$ fluxes $f_1, f_2$ on the world-volume of each brane in the two
stacks.  These stacks carry unbroken components $U(N_1)$, $U(N_2)$ of
the gauge group.  The $3$ and $3'$ stacks with $2M$ branes have no background
flux and their gauge group is $SO(2M)$.  All three stacks and their
images are on top of each other and have been separated for clarity.}
\label{fig1}
\end{figure}

We have restricted ourselves to smooth K3 surfaces since we are working 
in the supergravity limit. When the lattice spanned by the flux vectors 
contains the root lattice of an A-D-E gauge group, the K3 surface 
develops the corresponding A-D-E singularity \cite{Asp96a}.  The root lattice of an 
A-D-E group is generated by norm-squared $-2$ vectors; as discussed earlier, the 
supersymmetry conditions \eqref{eq:SUSY-conditions} require the 
corresponding rational curves to zero volume, resulting in a singular 
K3 surface. When one of the flux vectors $f_a$ satisfies $f_a \cdot f_a 
= -2$ the zero-volume rational curve also supports $N_a$ $U(1)$ instanton
fluxes.  In the
absence of instanton flux, the classical A-D-E singularity is smoothed
out by a strongly coupled worldsheet theory \cite{witten-ade}.  In the
absence of a singularity, $k$ coincident small instantons enhance the
gauge group by a factor $Sp(k)$, as shown in \cite{smallinstantons}.
When small instantons coincide with an A-D-E singularity, it was shown
in \cite{Aspinwall, intriligator, Blum1, Blum2, aspinwall-morrison}
that the coalescence of sufficiently many instantons can lead to a
further enhancement of the gauge group, with additional hyper and
tensor multiplets.  In our case, $N_a$ small $U(1)$ instantons with
$f^2 = -2$ carry the same six-form charge as $2N_a$ D5 branes, and
therefore there may be a gauge enhancement of at least $Sp(2N_a)$, with
further enhancement possible from the A-D-E singularity. It is not clear 
whether these singular compactifications are consistent string vacua, 
and we leave the further study of these singular models to future work.

The matter content is easily visualized in the type I picture, as
shown in Figure \ref{fig1} where we have depicted a two-stack model.
The gauge group is $U(N_1)\times U(N_2) \times SO(2M)$ with
$N_1+N_2+M=16$ and $M\geq 4$.  We have two stacks containing $N_1$ and
$N_2$ D9-branes, with distinct $U(1)$ fluxes characterizing the two
stacks.  The remaining $M$ branes do not have any flux on their
world-volume.  We denote the stacks and their orientifold images as 1,
2, 1', 2' and 3,3' respectively in Figure \ref{fig1}.  The strings
that stretch between $1\leftrightarrow 1$ and $2 \leftrightarrow 2$
correspond to the vector multiplet and transform in the adjoint of
$U(N_1)$ and $U(N_2)$ respectively.  The $1 \rightarrow 2$ strings
(and their CPT conjugate $2\rightarrow 1$ strings) are charged in the
$(N_1,\nb_2)+(\nb_1,N_2)$ representation of $U(N_1)\times U(N_2)$ .
There are also strings that stretch from a brane to the orientifold
image of another brane.  The $1\leftrightarrow 2'$ strings transform
in the bifundamental $(N_1,N_2)+(\nb_1,\nb_2)$.  The strings in the
$1\leftrightarrow 1'$ transform in the two-index antisymmetric
representation of $U(N_1)$.  The $1\leftrightarrow 2M$ and $2
\leftrightarrow 2M$ strings transform in the $(N_1,2M)+(\nb_1,2M)$ and
the $(N_2,2M)+(\nb_2,2M)$ representations respectively.

The massless matter content in these models can be computed using the
Atiyah-Singer index theorem \cite{gswest}, \cite{honecker},
\cite{honeckertrap}.  For example in the $1\leftrightarrow 2$ sector,
the ends of the string carry charge $(+\lambda,-\lambda)$ under the $U(1)\otimes
U(1)$ bundle $\mathcal{V}$ associated with the two branes to which the string is
attached ($\lambda$ determines the normalization of the Lie algebra generators 
and is defined in section 3.2).  The index theorem \cite{eguchi} gives the net number of
normalizable, fermion zero modes in the presence of this background
gauge field.
\begin{eqnarray}
n_+-n_- &  = & -\frac{1}{24} p_1(R) - c_2(\mathcal{V})  =  2 + \frac{1}{16\pi^2} \int \mathrm{tr}\, \CF \wedge \CF  \nonumber \\
& = &  2+ \frac{1}{2}(f_1-f_2)^2 \\
\Rightarrow |n_+-n_-| & = & -2-\frac{1}{2}(f_1-f_2)^2
\end{eqnarray}
Therefore, there are $(-2-(f_1-f_2)^2/2)$ 6D hypermultiplets in the $(N_1,\bar{N}_2)$ representation. 
This number is positive only when $(f_1-f_2)^2\leq -4$, which is consistent with the smoothness condition on $S$, {\it i.e.~}there are no integral norm-squared $-2$ vectors. 
The number of massless multiplets in the other representations can be
computed in an analogous manner.  For a more general $\prod_{a=1}^{k}
U(N_a) \times SO(2M)$ model, one can compute the above index for every
pair of stacks $a,b$; the results of this computation are shown in
Table \ref{tablematter}.  In addition, the closed string sector gives
the six-dimensional SUGRA multiplet, 1 tensor multiplet and 20 neutral
hypermultiplets \cite{polchi2}.  The important thing to note is that
{\it all} the massless matter content
is determined by inner products of the
lattice vectors $\{f_a\}$.    Defining the {\it intersection matrix}
$m_{ab}:=f_a\cdot f_b$,  the low-energy spectrum is thus completely
determined by $m$, with the gauge group determined by the stack sizes
$\{N_a\}$. It is interesting to note that requiring the number of multiplets
as determined by the matrix $m$ to be non-negative imposes a weak form of the smoothness criterion on $S$: the vectors $f_a$, $f_a\pm f_b$ must have norm-squared  $\leq -4$.

\begin{table}
\centering
\begin{tabular}{|c|c|}
\hline
Representation & Number of hypermultiplets \\
\hline
$(N_a,\nb_b)+(\nb_a,N_b)$ & $(-2-\frac{1}{2}(f_a-f_b)^2)$ \\
$(N_a,N_b)+(\nb_a,\nb_b)$ & $(-2-\frac{1}{2}(f_a+f_b)^2)$ \\
Antisym.  $U(N_a)$ + c.c & $(-2-2f_a^2)$ \\
$(N_a,2M)+(\nb_a,2M)$ & $(-2-\frac{1}{2}f_a^2)$ \\
Neutral &     20 \\
\hline
\end{tabular}
\caption{Massless multiplets in the $\prod_{a=1}^{k} U(N_a) \times SO(2M)$ solution.  The indices $a$ and $b$ run over stacks with $a\neq b$.} \label{tablematter}
\label{t:multiplets}
\end{table}

\subsection{Anomaly cancellation}

The anomalies in supersymmetric 6-dimensional theories can be cancelled 
by a generalization of the Green-Schwarz mechanism, as long as the $\Tr R^4$ 
and $\Tr F^4$ terms in the anomaly 8-form vanish. This computation was first 
carried out in \cite{gswest}, and more  recently in \cite{intriligator, 
honecker, honeckertrap}. The coefficient of the
$\Tr R^4$ term, which corresponds to the purely gravitational anomaly\footnote{
In 6D theories, the CPT conjugate of a spinor has the same chirality
as the spinor.  As a result when a 6D field theory is coupled to
gravity, in general there are purely gravitational anomalies \cite{gsw2}.}, is proportional to 
 $n_H-n_V+29n_T-273$, where $n_H$,$n_V$, $n_T$ are the
number of hypermultiplets, vector multiplets and tensor multiplets
respectively \cite{polchi2}. 
For the compactifications we consider,
the closed string sector fixes $n_T=1$ and contributes 20
gauge-neutral hypermultiplets (from the K3 moduli).  The gravitational 
anomalies vanish when $n_H-n_V=224$, if we only count 
the multiplets from the field theory.
It is easily checked that all our models satisfy this anomaly
constraint. 

Consider $K$ stacks, each of multiplicity $N_a$ and flux vectors $f_a$ 
satisfying the tadpole condition $\sum N_af_a^2=T=-48$.  The gauge group is $\prod_{a=1}^{K} U(N_a) \times SO(2M)$ with $\sum_{a=1}^K N_a + M = 16$.  The matter content is summarized in Table \ref{tablematter}.  The number of vector multiplets is $N_a^2$ for each $U(N_a)$ factor and $M(2M-1)$ for the $SO(2M)$.
\begin{equation}
n_V = \sum_{a=1}^K N_a^2 + M(2M-1)
\end{equation}
To obtain the number of hypermultiplets, we simply multiply the numbers in Table \ref{tablematter} with the dimension of the representation.  This gives
\begin{eqnarray*}
n_H & = & -\sum_{a\neq b} N_a N_b(2+\frac{1}{2}f_a^2+\frac{1}{2}f_b^2) - \sum_a N_a(N_a-1)(1+f_a^2)-2M\sum_a N_a(2+\frac{1}{2}f_a^2) \\
& = & -\sum_{a, b} N_a N_b(2+\frac{1}{2}f_a^2+\frac{1}{2}f_b^2)  +\sum_a N_a^2 + \sum_a N_a(1+f_a^2)-2M\sum_a N_a(2+\frac{1}{2}f_a^2) \\
& = & -2(16-M)^2-(16-M)T + \sum_a N_a^2+ 16-M+T-4M(16-M) - MT \\
& = & -496-15T+\sum_aN_a^2+2M^2-M \\
& = & n_V + 224
\end{eqnarray*}
Thus, all the models we construct satisfy the condition $n_H-n_V=224$ and are anomaly free.

\section{Existence of solutions from lattice embeddings}
\label{sec:theory}

We have now given a simple mathematical characterization of abelian
magnetized brane models on K3.  These models are specified by a set of
lattice vectors $\{f_a\}$, not necessarily linearly independent, which 
along with $\fwvec$ generate a negative-definite sublattice
$\M$ in $H^2(S,\Z)\cong \lat$.  Each lattice vector $f_a$ is
associated with $N_a$ identical abelian fluxes on a stack of $N_a$
D9-branes.  These lattice vectors and multiplicities satisfy the
tadpole condition 
\begin{equation}
{\rm Tr}\; \hat{m} =
\sum_a N_a f_a^2 = -48\,.
\label{eq:tadpole-theory}
\end{equation}
As shown in section
\ref{matter}, the gauge group and matter content of the resulting
6-dimensional supersymmetric theory depend only on the multiplicities
$N_a$ and the intersection matrix $m_{ab}:=f_a\cdot f_b$.

We are now interested in finding a simple way of classifying all
models of this type.  In particular, in order to study the physics of
this class of string compactification, we would like to know the
answers to the following two questions:
\vspace*{0.05in}

{\bf (I)} What combinations of multiplicities $N_a$
and intersection matrices $m_{ab}$ can be realized in string theory?
\vspace*{0.05in}

{\bf  (II)} For those $N_a$ and $m_{ab}$ which can be realized, in how
many inequivalent ways can models with this data be constructed?
\vspace*{0.05in}

The equivalences referred to here are changes of basis of the lattice
$H^2(S,\Z)$, which correspond to (large) diffeomorphisms of the K3
surface\footnote[5]{This is the Torelli Theorem for K3 surfaces
\cite{Barth}.  There is a simple analogy with the case of the 2-torus.
$H^1(T^2,\Z)$ is a lattice with an anti-symmetric bilinear form
$\left(\begin{array}{cc}0 & 1 \\ -1 & 0 \end{array}\right)$.
Automorphisms of this lattice form the group $SL(2,\Z)$, which is the
group of large diffeomorphisms of $T^2$. See also related
discussions
in \cite{Asp96a, Tripathy-Trivedi}.}.  
The questions above essentially amount to whether the lattice defined by
the intersection matrix $m_{ab}$\footnote[9]{The vectors $f_a$
 need not be linearly independent, therefore, the matrix $m_{ab}$ does define a lattice, but is not the {\it
inner product matrix} associated with the lattice $\M$.  This turns
out to be a minor subtlety and we address this issue in more detail in
Section \ref{sec:enumeration}.}  can be embedded in the 2-cohomology
lattice of K3, and whether such an embedding is unique.  We begin with 
the simple
case where there is only a single nonvanishing flux vector $f_a$, then
proceed to the more complex case of many nonvanishing flux vectors.
In this section we focus on addressing question (I), the existence of
models with given $N_a, m_{ab}$.  In the following section we discuss
question (II) and the explicit enumeration of models.

\subsection{Single Stack Models} 
\label{sec:single-embedding}

To illustrate the character of the existence and uniqueness results
for lattice embeddings which we will make use of,
it is  helpful to begin with the
simplest case, that of a single nonvanishing flux vector $f \neq
0$.  Since $f$ is in integral cohomology of K3, we have 
\begin{equation}
f \cdot
f = -\tau \in 2\Z \,.
\end{equation}
We can think of $f$ as defining a one-dimensional lattice.  This
lattice is negative-definite since $f \cdot f = -\tau < 0$.
Ignoring for the moment any physical constraints, the mathematical
question we wish to address is whether, for a given $\tau$, there
exists a vector in $H^2 (K3,\Z) = \lat$ which squares to
$-\tau$, and whether such a vector is unique under automorphisms of
the lattice ({\it i.e.} linear transformations on the lattice leaving
the inner product unchanged).

To understand the nature of this question it is perhaps helpful to
consider briefly a simpler version of this question, namely the
analogous question of existence and uniqueness of a vector of fixed
length but on a lattice of definite signature.  For example,
consider the 2D Cartesian lattice with generators $a_1, a_2$ having
inner product $a_i \cdot a_j = \delta_{ij}$.  There only exist vectors
$v = xa_1 + ya_2$ of norm-squared $v \cdot v = \nu$ when $\nu$ can be
written as a sum of squares $\nu = x^2 + y^2$ for integral $x, y$.
Thus, for example $\nu =5 = 1 + 4$ is associated with vectors such as
$a_1 + 2 a_2$, while for values such as $\nu = 6$ which
cannot be written as a sum
of squares there are no vectors in this lattice with norm
$\nu$.  On this lattice, there are sometimes multiple vectors of
the same length which are not related by automorphisms (for example 65
= 1 + 64 = 16 + 49).  The number of distinct automorphism classes of
vectors of length squared $\nu$ is the number of ways $\nu$ can be
expressed as a sum of squares (for which a formula was found by Gauss).
Thus, on this lattice, we are not guaranteed the existence of a vector
with some arbitrary value $\nu$ for the norm.  And, even if a vector of
norm-squared $\nu$ exists in the lattice, we are not guaranteed that it is unique.

For an even unimodular lattice such as $E_8$, the existence question
has a simpler answer --- there are vectors in $E_8$ with any desired
even norm-squared (See Appendix \ref{app_lat}).  As on the Cartesian 2D lattice,
however, there can be multiple vectors of fixed length in different
orbits of the automorphism group (for example there are two disjoint
orbits containing vectors of norm-squared 8, and two orbits with vectors of
norm-squared 14).  The number of such orbits for fixed $\nu$ can be found by
looking at vectors in a fundamental domain for the action of the
automorphism group; for $E_8$, for example, these numbers are given in
 \cite{e8}.

For lattices with indefinite signature, the uniqueness problem also
simplifies dramatically.  The simplest example of this is the lattice
$U \oplus U$ defined by two copies of the basic indefinite signature
matrix
\begin{equation}
U =\left(\begin{array}{cc} 0 & 1\\1 & 0\end{array}\right).
\end{equation}
On this lattice, as generally occurs for lattices of indefinite
signature, automorphism classes of vectors are essentially uniquely
defined by the vector norm.  To make this statement more precise, it
is useful to define a {\it primitive} vector $v$ in a lattice $\L$ to
be a vector with the property that $v$ cannot be written as an
integral multiple $v = dw$ of another vector $w \in \L$.  If $v = dw$
where $w$ is primitive and $d > 1$ is integral, then $v$ is not
primitive, and $d$ is the {\it divisor} of $v$.  The divisor is
invariant under lattice automorphisms; the orbit of a primitive vector
under the automorphism group does not contain any non-primitive
vectors.  For example, of the two automorphism classes of vectors of
norm-squared 8 in $E_8$, one contains a primitive vector of norm-squared 8 and the
other contains $2v_2$ where $v_2$ is a primitive vector of norm-squared 2.
Given this definition of primitive vectors, it can be proven by
elementary arguments that any two primitive vectors in the lattice $U
\oplus U$ with the same norm-squared are related by a lattice automorphism.
This result is shown by Wall in \cite{wall}, and generalized by
induction to any lattice whose net signature (number of positive
eigenvalues minus number of negative eigenvalues) differs from the
rank by at least 4.  Moreover, indefinite, even, unimodular lattices
like $\lat$ contain primitive vectors of any even norm, since they
always contain factors of $U$\footnote[5]{If $v$ and $w$ are vectors
spanning the lattice $U$ with $v^2 = w^2, v \cdot w = 1$, then for a given even integer $2k$, the
primitive vector $v+kw$ has norm-squared $2k$.  Since any indefinite, even,
unimodular lattice contains factors of $U$ by Milnor's classification
(See Appendix \ref{app_lat}), such a lattice contains vectors of
arbitrary even norm.}.  Therefore, given any even integer $\nu$, there
exists precisely one primitive vector of norm-squared $\nu$ up to automorphism
in $\lat$.  This can be easily generalized to the case of
non-primitive vectors.  If $f$ is a non-primitive vector of norm
$\nu$, then $f=d g$, where $g$ is a primitive vector of norm-squared $\nu/d^2
\in 2\Z$.   Wall's theorem states, then, that there is a unique $g$ of
norm-squared $\nu/d^2$ as long as $\nu/d^2$ is even.  To summarize, we have
the result that

\begin{center} {\it
Equivalence classes of vectors under automorphisms in $\lat$ are
uniquely identified by norm-squared and divisor; there is exactly one equivalence class for any even norm-squared $\nu$ and divisor $d$ satisfying $(2d^2) \ | \ \nu$.
}
\end{center}

Bringing this result back to the context of line bundles on K3, this
states that a line bundle with flux $f$ can be constructed with $f^2 =
-\tau$ for any even integer $\tau$.  Equivalence classes of such line
bundles are determined by $\tau$ and a divisor $d$ such that $d^2 |
\tau/2$.  
This gives a simple and complete classification of possible supersymmetric line
bundles on K3.  Including the tadpole constraint restricts $N f^2 = -N
\tau = -48$, so there are in practice only a small number of allowed
configurations.  We discuss explicit enumeration of solutions in
Section \ref{sec:enumeration}.

\subsection{Multiple stack models}
\label{sec:multiple-stack-models}

Now let us consider the more general situation of $K$ stacks of
fluxes, where the $a$th stack contains $N_a$ D9-branes with flux
$f_a$.  The intersection matrix $m_{ab} = f_a \cdot f_b$ defines a
negative-definite lattice of dimension $\leq K$.  When the $f_a$ are
linearly independent, the matrix $m$ is negative-definite, and the
dimension of the lattice is $K$.  The matrix $m$ can be degenerate;
this occurs when the fluxes $f_a$ are not linearly independent.  In
this case $m$ still defines a lattice, but of dimensionality $< K$.
We discuss the situation where the $f_a$ are not linearly independent
in more detail in the following section.
We are now interested in answering questions (I) and (II) above for
negative-semidefinite matrices $m_{ab}$ corresponding to
negative-definite lattices of  dimension $\leq 12$.

In principle, classifying allowed matrices $m_{ab}$ seems like a very
difficult question as the dimensionality of the matrix increases.  One
might be tempted to begin by classifying the lattices associated with
allowable intersection matrices (up to automorphism), which correspond
to integral quadratic forms, and then ask if the resulting lattices
can be embedded into $\Gamma^{3, 19}$.  The classification of integral
quadratic forms is a classical problem in mathematics, with a long
history.  Gauss classified all binary (two-dimensional) integral
quadratic forms in {\it Disquisitiones Arithmeticae} \cite{Gauss}.
Later work by Minkowski, Hasse, Witt, Eichler and others has extended
this work to higher-dimensional and indefinite-signature forms.  Even
using powerful $p$-adic methods, however, the classification of
integral quadratic forms becomes intractable beyond a certain point.
A clear review of this work is given in \cite{Conway-Sloane}.  While
 the integral quadratic forms of dimension up to 12, which
would be of interest to us here, can in principle be classified, this is 
a cumbersome way to analyze the problem with which we are faced.

Since the data associated with a given line bundle model
provides not only the abstract lattice (integral quadratic form)
defined by $m_{ab}$, but also a choice of vectors $f_a$ in this
lattice, we can take a clearer path to the classification of allowed
models.  In doing so we use a generalization by Nikulin of Wall's 
theorem (see previous section) to higher dimensions 
\cite{nikulin1} and proceed in analogy with the one-stack case.
In this section we show
how these theorems apply, and discuss the limiting cases of their
application.  In Section \ref{sec:enumeration} we show more explicitly
how to classify and enumerate explicit models with desired features
based on the results of the more theoretical analysis contained in
this section.

To understand the statement of Nikulin's generalization of Wall's
result, we need to generalize the notion of primitivity to a lattice
embedding of higher dimensionality.  Given a lattice $\M$ and another
lattice $\L$, an embedding $\phi:\M \rightarrow \L$ is {\it primitive} if
for all primitive vectors $x \in \M, \ \phi (x)$ is primitive in $\L$.
Basically, an embedding is primitive if the image of $\M$ in $\L$
contains all vectors in $\L$ in the $\R$-linear space spanned by $\phi
(\M)$.  A slightly weakened version of the theorem of Nikulin 
is almost completely adequate for our purposes.
\begin{theorem}[Nikulin, simplified] \label{embed2}
Let $\M$ be an even lattice of signature $(t_+,t_-)$ and let $\L$ be an
even, unimodular lattice of signature $(l_+,l_-)$.  There exists a
primitive embedding of $\M$ into $\L$
which is unique up to automorphisms of $\L$, provided the following
conditions hold:
\begin{enumerate}
\item $l_+-t_+ > 0$ and $l_--t_->0$.
\item $l_++l_--2t_+-2t_- \geq 2$
\end{enumerate}
\end{theorem}
This is a simplified version of Theorem 1.14.4 in \cite{nikulin1}.
The full theorem is slightly stronger and has conditions depending on
prime components of the abelian group $\M^*/\M$, where
$\M^*:=\mbox{Hom}(\M,\Z)$ denotes the dual lattice of $\M$.  In
Appendix \ref{app_embed} we give a more precise statement of Nikulin's
stronger theorem, proven by Nikulin using $p$-adic methods, and show
how Theorem \ref{embed2} follows from Nikulin's theorem.  For a
further discussion of embedding theorems, see \cite{dolgachev, 
morrison}.  These embedding theorems are also encountered in a
physics context in \cite{Moore}.

From Theorem \ref{embed2}, it follows immediately that any even,
negative-definite lattice $\M$ of dimension up to 10 has a primitive
embedding in the lattice $\L = \Gamma^{3, 19}$.  Thus, Nikulin's
theorem guarantees that any intersection matrix describing up to 10
linearly independent abelian fluxes 
can be realized on a sublattice of the cohomology
lattice of K3.  In addition, the theorem states that there is a {\it
unique} primitive embedding  of any such sublattice into $\lattice$
up to automorphism.  This is completely
analogous to the one-dimensional case where Wall's theorem states that
there is a unique primitive vector for a given norm.

From the tadpole constraint $\sum_{a}N_af_a^2 = -48$ and the fact that
$f_a^2 \leq -4$, we can have at most 12 non-zero fluxes, corresponding
to a maximum lattice rank of 12.  So Nikulin's theorem almost
completely covers the cases of interest.  We cannot draw general conclusions
when the rank of the lattice is $11$ or $12$, and these have to be analyzed
on a case-by-case basis. For this purpose, we state Nikulin's theorem in 
greater generality in Appendix \ref{app_embed}.

The upshot of this analysis is that any
negative-semidefinite even intersection matrix $m_{ab}$ of rank $\leq 10$ gives a
negative-definite lattice which admits a primitive embedding into
$\Gamma^{3, 19}$, and this embedding is unique up to automorphisms of
$\Gamma^{3, 19}$.  Physically, this means that the only constraint on
allowed intersection matrices is the tadpole constraint, so that up to
this constraint any desired model can be realized by dialing the
intersection matrix.  Furthermore, for primitive embeddings the
resulting realization is unique up to 
automorphisms of the K3 cohomology lattice.  This result could break down
when the rank of the intersection matrix $m$ equals 11 or 12.

In the analysis of this section we have concentrated on primitive
embeddings.  In the general classification of models we must include
non-primitive embeddings, which leads in some cases to a discrete
degeneracy of models with the same gauge group and matter content.  
This leads to the following result --
\begin{center}
{\it Every negative-semidefinite intersection matrix $m$ of rank $\leq 10$ that is even
and satisfies the tadpole constraint gives a lattice $\M$ which has an
embedding (primitive or otherwise) into $\lat$.}
\end{center}
In the next section, we use the general existence and uniqueness
result described above to provide explicit methods for classifying and
enumerating models, including discrete degeneracies arising from
non-primitive embeddings.  We also describe in more detail the case
when the intersection matrix $m$ is negative-semidefinite but not
negative-definite, corresponding to the case when the vectors $f_a$
are linearly dependent.

\section{Classification and enumeration of vacuum solutions}
\label{sec:enumeration}

Given the mathematical results on lattice embeddings described in the
last section, we are now equipped to give a simple description of the
full set of possible vacuum solutions.  Each vacuum solution is
associated with an even, negative-semidefinite intersection matrix
$m_{ab}$, associated with a negative-definite lattice $\M$, as well as
multiplicities $N_a$.  The low-energy gauge group and matter content
of the dimensionally reduced 6D theory depend only on $m_{ab}, N_a$.
To enumerate all solutions we consider all possible
negative-semidefinite intersection matrices and multiplicities, 
satisfying the tadpole constraint.  Each such model admits a unique
primitive embedding into $\Gamma^{3, 19}$ (except possibly for some cases when the rank of $m$
is
11 or 12).  In some cases, non-primitive embeddings can also be
constructed.  In this section we show how non-primitive embeddings can
be constructed, associated with refinements (overlattices) of the
lattice defined by $m$.  This gives a direct approach to construction
of all models associated with given data $m_{ab}, N_a$.

\subsection{Single stack models}
\label{sec:single-enumeration}

The simplest case of the magnetized brane construction on K3 we are
considering here is a single stack of $N$ D9-branes carrying a flux
$f$ (rank $m = 1$).  The fluxes of the $U(1)^{16}$ subgroup of $SO(32)$ are then
$f_1=f_2=\cdots=f_{N}=f, f_{N+1}=\cdots=f_{16}=0$.  The 6D  gauge group
of this model is $U(N)\times SO(32-2 N)$, and the matter content
depends only on the quantities $N$ and $f$.
In this single stack model,
$N$ and $f$
satisfy the tadpole constraint $Nf^2=-48$.  In addition, we require the vector $Nf/2$
to be integral. When $N$ is even, the one-dimensional lattice $\M$ is generated by $f$,
and when $N$ is odd, $\M$ is generated by $f/2$. 

The data describing this model are just the integer $N$ and the
(even) integer $\tau = -f^2 \geq 4$.  To classify models of this type we need
to find all pairs $\tau, N$ satisfying the above constraints.  We must then
find all embeddings of the one-dimensional lattice $\M$ into $\Gamma^{3, 19}$.

The solutions are
\begin{equation}
(N,\tau) = (1,48), \ (2,24), \ (3,16), \ (4,12), \ (6,8), \ (8,6), \ (12,4)\, . \label{eq:tau1}
\end{equation}
Let $v$ denote the generator of the lattice $\M$, which is $f$ 
for even $N$ and $f/2$ for odd $N$. We see that $-v^2$ is an
even integer $\geq \ 4$. From Wall's theorem quoted in \ref{sec:single-embedding}, there exists a
primitive vector of norm-squared $v^2\, $ in $\, \Gamma^{3, 19}$, and this vector
is unique up to automorphisms.  In other words, the one-dimensional lattice $\M$
admits a unique primitive embedding into $\lat$.
Thus, there is at least one model
associated with the data $(N, \tau)$ for each choice satisfying the
tadpole constraint.

The only remaining question is for which values of $\tau$ the
one-dimensional lattice defined by $v$ admits a non-primitive
embedding.  In the one-dimensional case this is a rather trivial
problem.  For a non-primitive embedding, we must have $v = d\, v'$, where
the divisor $d$ is an integer and $v'$ is embedded primitively.  
So we
simply need to identify all integers $d$ whose square divides
$v^2/2$.  Also, the lattice spanned by $v'$ must not contain any $-2$ vectors. 
For each such integer there is a unique primitive embedding
of the associated vector $v'$ of norm-squared ${v}^2/d^2$.  The only solution in
(\ref{eq:tau1}) where this is possible is 
$(N,\tau) = (2,24)$, where $v = 2v'$.

This analysis has thus allowed us to classify the topologically
distinct magnetized brane models on a K3 surface with a single stack of
identical abelian fluxes, where the tadpole cancellation
condition is satisfied.  For all these models, since $f^2 < 0$ there
exists a parameter space of values for $\Omega$ and $J$ that preserve
supersymmetry.  We summarize the possible vacua in Table \ref{table3},
indicating the gauge group and matter content of the 6D theory in each
case, along with the number of topologically distinct ways of
realizing each model.

\begin{table}
\centering
\begin{tabular}{|c|c|c|c|c|}
\hline 
& & Antisym.  & &\\
$\tau = -f^2 $ &Gauge group &  $U(N)$+cc & $(N,2M)+ cc$ & \# vacua \\
\hline 
48 &$U(1) \times SO(30)$ & 0  & 22 & 1\\
24 &$U(2)\times SO(28)$ & 46 & 10 & 2\\
16 &$U(3)\times SO(26)$ & 30 & 6 & 1\\
12 &$U(4)\times SO(24)$ & 22 & 4  & 1\\
8 &$U(6)\times SO(20)$ & 14 & 2  & 1\\
6 &$U(8)\times SO(20)$ & 10 & 1  & 1\\
4 &$U(12)\times SO(8)$ & 6  & 0  & 1\\
\hline
\end{tabular}
\caption{Solutions with a single $U(1)$ flux.  The third and fourth
columns give the number of hypermultiplets in the representation
indicated.  The last column gives the number of (topologically
distinct) choices of flux that yield the corresponding low-energy
theory. }\label{table3}
\end{table}

Note that the low-energy theories associated with the topologically
distinct models where $f$ admits a non-primitive embedding are
identical at the level of the gauge group and matter content with
those models where $f$ is embedded primitively, although
there is a discrete topological quantum number associated with the
form of the embedding in $\Gamma^{3, 19}$ that distinguishes these
models.  This general pattern is reproduced for models with more
linearly independent fluxes $f$ (more stacks).  It would be
interesting to consider precisely how these discrete sets of models
with identical gauge group and matter content are distinguished by
more detailed aspects of the six-dimensional physics.

\subsection{Multi-stack models}
\label{sec:multi-stack-models}

Now let us consider the general situation where there are multiple
distinct stacks of branes ($K > 1$) with different fluxes $f_a, a = 1,
\ldots, K$ and multiplicities $N_a$.  The intersection matrix $m_{ab}
= f_a \cdot f_b$ defines a lattice $\M \subset\lattice$.  We are
interested in classifying all such configurations.  We can do this
using the lattice embedding theorems discussed in \ref{sec:theory}.
To apply these theorems, however, there are two technical issues that
must be addressed.  
First, we need to deal with the situation where the lattice $\M$ is not
embedded in a primitive fashion into $\lattice$, generalizing the
discussion of divisors in the previous subsection.  
Second, we need to deal with the fact that the
$f_a$ need not be linearly independent, leading to cases where the
matrix $m$ is negative-semidefinite, not negative-definite.  

\subsubsection{Primitive embeddings and overlattices}
\label{sec:primitive-over}

Let us first consider the question of primitivity.  We deal with the
case of linear dependencies between the $f$'s in the following
subsection \ref{sec:dependencies}.  We assume in the analysis of this
subsection that the vectors $f_a$ are linearly independent, so that
the matrix $m$ is negative-definite.  As we will demonstrate in the
next subsection, essentially the same analysis will work in the
degenerate case, by working with a linearly independent subset of the
$f$'s.  

We know that the lattice $\M$ defined by $m$ admits a unique (up to
automorphism) embedding into $\lattice$ (with the usual caveat that
rank $m \neq 11, 12$).  Thus, for every negative-definite $m$ satisfying the tadpole
condition there is at least one string model realizing this
intersection matrix.  In some cases, however, there are other models
that realize the same intersection matrix $m$ through a non-primitive
embedding of $\M$ into $\lattice$.  Thus, to provide a complete
classification of models we must understand the range of possible
non-primitive embeddings of any lattice.

An embedding $\M \subset\lattice$ fails to be primitive when there are
lattice points in $\lattice$ that lie in the $\R$-linear subspace spanned
by $\M$ but not in $\M$ itself. In other words, the real plane in $\lat$ that contains the lattice $\M$, also contains other lattice points that are not in $\M$. 
A lattice $\N$ is said to be an {\it overlattice} of a lattice $\M$, if $\M$ is a proper sub-lattice of $\N$ and if both $\M$ and $\N$ have the same dimension. 
When the embedding $\M \hookrightarrow \lat$ is non-primitive, the set of lattice
points in $\lattice \cap {\rm span} (\M)$ forms an {\it overlattice}
of $\M$, {\em i.e.} a more dense lattice containing $\M$ as a
sublattice of equal dimension. The lattice spanned by $f' = f/d$ in
the case of a non-primitive single stack model discussed above is a
simple example of an overlattice.

\begin{figure}
\centering
\includegraphics[width=\textwidth]{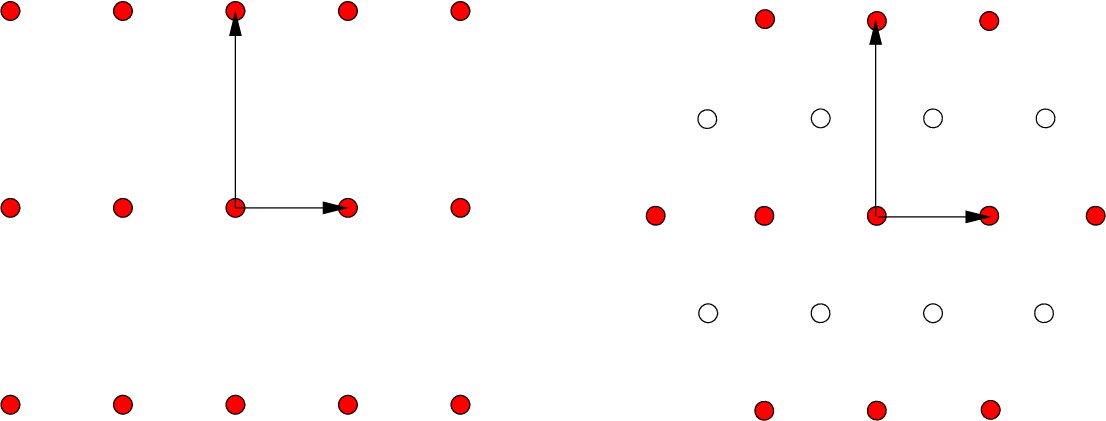}
\caption{A lattice $\M$ is shown on the left, and its overlattice $\N$ on the right.}
\label{fig:sec6lattice}
\end{figure}

For any lattice $\M$, which is generated by $K$ linearly independent
vectors $f_a$, we can enumerate all possible embeddings $\M
\rightarrow\lattice$ by enumerating all (even) overlattices
$\M'\supset\M$.  By the lattice embedding theorems, there is then a
unique (up to automorphism) non-primitive embedding of $\M$ for each
allowed $\M'$, associated with a primitive embedding of $\M'$.

Classifying overlattices of a given lattice $\M$ with basis $f$ can be
done in a straightforward fashion by induction.  Assume $\M$ is
generated by linearly independent lattice vectors $f_1, \ldots, f_K$.
Given any overlattice $\M' \supset\M$, we can choose a basis $e_1,
\ldots, e_K$
for $\M'$ inductively so that $e_1, \ldots, e_a$ form a basis for the
sublattice of $\M'$ spanning the space containing $f_1, \ldots, f_a$.  We then have
\begin{eqnarray}
f_1 & = &  \alpha_{11}e_1\nonumber \\
f_2 & = &  \alpha_{21} e_1 + \alpha_{22} e_2 \label{eq:over-linear} \\
\vdots & & \vdots \nonumber \\
f_K & = & \alpha_{K1} e_1 + \cdots + \alpha_{KK} e_K \, .  \nonumber
\end{eqnarray}
Geometrically, at each step of
the construction $e_a$ is a vector chosen to have minimal extent in
the direction defined by the component of $f_a$ perpendicular to the space spanned by
$\{f_1, \cdots, f_{a-1}\}$.
Defining the inner product matrix on the $e_a$'s by $\epsilon_{ab} =
e_a \cdot e_b$, we thus have for each $a \leq K$
\begin{equation}
\det_a m = \prod_{b = 1}^a \alpha_{bb}^2 \ \det_a \epsilon \,.
\label{eq:diagonal-product}
\end{equation}
where by $\det_a m$ we mean the determinant of the $a \times a$
principal minor of $m$.  
Note that when $a$ is not divisible by 8, $|\det_a\epsilon| > 1$, as there
are no negative-definite, even, unimodular lattices in dimensions not divisible by 8.
At each stage of the inductive definition of the $e_a$'s, the basis
vector $e_a$ can be shifted by a linear combination of $e_b$ with $b <
a$ so that the coefficients $\alpha_{ab}$
satisfy
\begin{equation}
0 \leq\alpha_{ab} < \alpha_{aa}, \;\;\;\;\; {\rm for} \; b < a \,.
\label{eq:a-inequality}
\end{equation}  
These inequalities fix the $SL(K,\Z)$ freedom associated with the choice 
of basis for the lattice $\M'$.

It is now straightforward to construct all overlattices of $\M$.  We
first enumerate all combinations of (even) diagonal elements $\alpha_{aa}$
satisfying (\ref{eq:diagonal-product}).  We then consider all
$\alpha_{ab} < \alpha_{aa}$.  The resulting matrix $\alpha$ then
defines an overlattice of $\M$ through (\ref{eq:over-linear}) if all
resulting inner products $\epsilon_{ab} = e_a \cdot e_b$ are integral.

This gives a systematic way of enumerating all overlattices of $\M$
for any even integral lattice.  For any given intersection matrix $m$,
there can be multiple distinct realizations of the corresponding
physics, labeled by matrices $\alpha$ satisfying
(\ref{eq:a-inequality}) and (\ref{eq:diagonal-product}).  The matrix
entries $\alpha$ then form a set of discrete quantum numbers labeling
the vacua.  This same structure arises in enumerating dyon states; a
similar analysis was performed by Banerjee and Sen in the context of
dyons in  \cite{Sen-dyon}, corresponding to the case of a
two-dimensional matrix $\alpha_{ab}, 1 \leq a, b \leq K = 2$.
The discrete quantum numbers $\alpha_{ab}$ characterizing distinct
vacua with the same gauge group and matter content associated with
coefficients in  (\ref{eq:over-linear}) can be thought of as a
generalization to higher dimension of the discrete invariants
identified in  \cite{Sen-dyon}.

\subsubsection{Fluxes with linear dependencies}
\label{sec:dependencies}

Now let us return to the degenerate case where $m$ has vanishing
determinant, associated with linear dependencies between the $f_a$'s.
We have already dealt with a simple class of such examples, namely
those where multiple $f_i$ are identical.  In this case we have
simplified the story by associating each set of branes with identical
fluxes with a single stack of $N_a$ branes with flux $f_a$.  To deal
with the more general case of linear dependencies, we can proceed by
working with a maximal subset of linearly independent $f_a$'s.  Enumerating
such a subset as $\tilde{f}_n, n = 1, \ldots, \tilde{K}$, we have a
basis for a $\tilde{K}$-dimensional real vector space. Each of the $f_a$'s not in the
linearly independent subset $\{\tilde{f}_n, n = 1, \ldots, \tilde{K}\}$ can be expressed as a linear
combination of $\tilde{f}_n$'s through relations of the form
\begin{equation}
f_a = \sum_{n}\gamma_{an} \tilde{f}_n
\label{eq:identifications}
\end{equation}
where the coefficients $\gamma_{an}$ are rational.

The complete set of vectors $f_a$ live in a lattice $\M$ that has
dimensionality $\tilde{K}$.  While the $\tilde{f}_n$'s may not form a
generating basis for $\M$, we can identify $\M$ as the set of points
given by the set of all integral linear combinations of $f_a$'s
subject to the identifications (\ref{eq:identifications}).  The
lattice $\M$ is thus clearly an overlattice of the lattice generated
by the $\tilde{f}_n$'s.  We can therefore identify all overlattices of
$\M$ by constructing all even integral
overlattices $\M'$ of $\oplus\Z \tilde{f}_n$.
For each such $\M'$ there is a basis $e_b$ of $\M'$ related to the
$\tilde{f}_n$'s by (\ref{eq:over-linear}).  $\M'$ is an overlattice of
$\M$ when all $f_a$'s are expressed in terms of integral linear
combinations of the $e_b$'s through the composition of
(\ref{eq:identifications}) and (\ref{eq:over-linear}).  In other
words, we have an overlattice of $\M$ (or $\M$ itself) when
\begin{equation}
f_a = \sum_{n, b}\gamma_{an} \alpha_{nb} e_b \in\M', \;\;\;\;\;
\forall a,
\label{eq:mol}
\end{equation}
which occurs iff $\sum_{n} \gamma_{an} \alpha_{nb} \in\Z$ for all $a,
b$.
\vspace*{0.1in} Since the volume of the unit cell (discriminant) of
$\M'$, given by $\det\ \epsilon$, divides that of $\M$, we can
identify $\M$ as the lattice $\M'$ of maximum discriminant satisfying
the conditions (\ref{eq:mol}).
\vspace*{0.1in}

\subsubsection{Systematic analysis of multi-stack models}
\label{sec:algorithm}

Combining the considerations in the preceding two subsections, we have
a systematic procedure for enumerating all multiple-stack magnetized
brane models on K3.  The steps in this procedure are

\begin{enumerate}
\item Consider all integer solutions to $\sum_{a} N_a m_{aa} = -48$,
  where $N_a$ are integer stack sizes, and $m_{aa}$ are even integers
  corresponding to $f_a^2$.

\item Scan over all integer matrix entries $m_{ab}$ so that the matrix
  $m$ is negative-semidefinite.  A simple test for a matrix to be
  negative-semidefinite is the {\it Sylvester criterion:} a matrix
  $m$ is
  negative-semidefinite iff all principal minors (square matrices of
  any size in the upper left corner) of $-m$ have non-negative determinants.
  An efficient test for a matrix to be negative-semidefinite is to
  perform the {\it Cholesky decomposition} of the matrix $m = -l
  l^{\rm T}$ where $l$ is lower-triangular with non-negative entries.
  This test can be carried out in order $K^3$
  operations\footnote[4]{Although a given $K\times K$ matrix can be
  tested for negative-semidefiniteness efficiently, we do not have a
  particularly efficient algorithm to enumerate {\it all}
  negative-semidefinite matrices with fixed diagonal elements up to
  permutation symmetries.}  \cite{numericalrecipes}.

\item Choose a minimal linearly independent set of fluxes
  $\tilde{f}_n$.  This can be done easily in combination with the
  Sylvester criterion in the previous step; choose fluxes one by one,
  throwing out those fluxes that when added to the previous set give
  a vanishing determinant.

\item Retain solutions in which $v = \fwvec$  can be an integral lattice vector. 
This is true if $v^2 \in 2\Z$, and $v\cdot \tilde{f}_n \in \Z$ for all $n$.

\item Construct all overlattices of $\oplus\Z \tilde{f}_n
\oplus\Z v$ by scanning
over solutions to (\ref{eq:diagonal-product}) and then
(\ref{eq:a-inequality}), testing for integral values for
$\epsilon_{ab}$ and (\ref{eq:mol}).  This determines the discrete
multiplicity with which the model associated with intersection matrix
$m$ arises.  

\end{enumerate}

This procedure will generate all allowed configurations including
those in which the lattice contains a vector of norm-squared  $-2$,
corresponding to a singular K3 surface.  While some models with such
vectors may have consistent string realizations we do not consider
them here, leaving further exploration of models corresponding to
singular geometries for further work.  The consistent models
associated with smooth K3 geometries will form a subset of those
models identified through the above procedure.
In general, identifying those lattices that do not contain
norm-squared  $-2$ vectors is tricky; it is well known that finding the
shortest vector in a lattice is a NP-hard problem.  For small rank of
$m$, however, this problem is tractable.

\subsection{Example: Two-stack models}
\label{sec:two-stack-models}

As a simple example of the above analysis, let us consider two-stack
models.  With two stacks corresponding to the lattice vectors $f_1$
and $f_2$, the  gauge group is $U(N_1)\times U(N_2)\times SO(32-2N_1
-2N_2)$.  
To construct all models of this type, we first find solutions to the
tadpole condition
\begin{equation}
N_1f_1^2+N_2f_2^2=-48 \,.
\label{eq:2-tadpole}
\end{equation}
For even $f_a^2$, solutions can only exist when
$\gcd(N_1, N_2) | 24$.
Each solution of (\ref{eq:2-tadpole})
gives diagonal elements in
the matrix $m_{ab}$ which for convenience we denote
\begin{equation}
m = \left(\begin{array}{cc} m_{11} & m_{12} \\ m_{12} &
m_{22}\end{array} \right) = \left(\begin{array}{cc} A & B \\ B &
C\end{array} \right) \label{eq:m2}
\end{equation}
We first consider the case when the matrix $m$ is non-degenerate.  In
other words, the vectors $f_1$ and $f_2$ are linearly independent.
Since $m$ is negative-definite, and we are only interested here in
lattices without vectors of self-intersection $-2$, we have $A\leq -4, C
\leq -4, AC-B^2>0$.  For each solution of (\ref{eq:2-tadpole}), we
thus need only consider varying $B$ in the range
$-\sqrt{AC} < B < \sqrt{AC}$.  This generates all
possible matrices $m$.  To consider only models from smooth K3's we
must further check that there are no $-2$ vectors in the lattice
generated by $f_1, f_2$.  While for matrices $m$ of arbitrary rank
this is a hard problem, for rank 2 there is a simple algorithm
(originally due to Gauss) for finding the shortest vector in a
lattice.  If $|m_{11} | > | m_{22} |$ and $2 | m_{12} | \leq | m_{22}
|$ then $f_2$ is the shortest vector.  If the second condition is not
satisfied then we replace $f_1$ with $\tilde{f} = f_1 + mf_2$ such
that $| \tilde{f} |^2$ is minimized, and then repeat the procedure,
exchanging $f_1, f_2$ as necessary, until we have the shortest vector,
of which we check the norm-squared to see if it is $-2$.  
As a final step, we only keep solutions where the vector $v = \fwvec$
is an integral lattice vector.
The range of possible
$N$'s and $m$'s produced in this fashion gives all possible gauge
groups and matter content for two-stack models.  To count the
topologically distinct ways in which one of these low-energy theories
can be realized in a string compactification, we have to count
possible overlattices of the lattice $\M$ associated with $m$.

For a given matrix of the form (\ref{eq:m2}), we can check for
overlattices as described above.  First, we consider all $\alpha_{11}$
so that $\alpha_{11}^2 | (A/2)$ (the extra factor of 2 arises from
(\ref{eq:diagonal-product}) because $\epsilon_{11}$ is even).  Then we
consider all $\alpha_{22}$ such that $\alpha_{22}^2 | (A B -C^2)/ \alpha_{11}^2$,
and finally all $\alpha_{21}$ in the range $0 \leq \alpha_{21} <
\alpha_{22}$.
We can now find expressions for each element $\epsilon_{ab}$ from
(\ref{eq:over-linear}), by computing the matrix elements $m_{ab} = f_a
\cdot f_b$  in terms of $\epsilon_{ab}$ and then solving for
$\epsilon_{ab}$.  In particular we have
\begin{eqnarray}
\epsilon_{11} & = &  \frac{m_{11}}{ \alpha_{11}^2}  \nonumber\\
\epsilon_{12} & = &  \frac{m_{12}-\alpha_{11} \alpha_{21}
  \epsilon_{11}}{ \alpha_{11} \alpha_{22}} \label{eq:e12}\\
\epsilon_{22} & = & \frac{m_{22}-\alpha_{21}^2 \epsilon_{11}
- 2 \alpha_{21} \alpha_{22} \epsilon_{12}}{ \alpha_{22}^2} 
\nonumber
\end{eqnarray}
Of those values for $\alpha_{21}$ in the allowed range, only those
giving integer values for $\epsilon_{12}$ correspond to integral
overlattices.
Again, to find models associated with smooth K3 compactifications we
must check each overlattice for a $-2$ vector as above.

\vspace*{0.1in} {\bf Example 1:} Let us consider the example of
solutions with $N_1 = N_2 = 6$, {\it i.e.}  gauge group $U(6) \times
U(6)\times SO(8)$.  The only solution to (\ref{eq:2-tadpole}) is $A =
-4, B = -4$.  So the matrix $m$ is given by
\begin{equation}
m = \left(\begin{array}{cc} -4 & B \\ B &
-4\end{array} \right) 
\end{equation}
where the allowed values for $B$ are $B = 0, \pm 1, \pm 2, \pm 3$.
For $B = \pm 3$ there is a vector ($f_1 \pm f_2$) of norm-squared  $-2$.
These are not good 6D gravity theories since the spectrum of one type
of bifundamental fields becomes negative.
Thus, for models with the  gauge group $U(6) \times
U(6)\times SO(8)$, there are 5 distinct allowed negative
semi-definite matrices providing theories with distinct matter content
in 6D.  It is easy to see that all these lattices are
negative-definite, and none admit overlattices.

\vspace*{0.1in}
{\bf Example 2:}  For an example where an overlattice is possible,
consider the solution of (\ref{eq:2-tadpole}) given by $N_1 = N_2 = 2$
with matrix
\begin{equation}
m = \left(\begin{array}{cc} -12 &  0 \\  0 &
-12\end{array} \right) 
\label{eq:sec6lattice}
\end{equation}
By the general theorem of Nikulin, the corresponding lattice $\M$
(shown in Figure \ref{fig:sec6lattice}) admits a primitive embedding into
$\lattice$.  This lattice also, however, admits an overlattice with a
primitive embedding, defining a distinct model with the same gauge
group and matter content.  To identify the overlattice we follow the
above procedure.  Again, $\alpha_{11} = 1$ since $A = -12$ and therefore
$f_1$ must be primitive.  The determinant is $\det\ m = 144$, which is
divisible by $\alpha_{22}^2$ for $\alpha_{22} = 2, 3, 4, 6, 12$.  
Choosing $\alpha_{22} = 2$, we then have from
(\ref{eq:e12}) $\epsilon_{12} = 6 \alpha_{21}$
and $\epsilon_{22} = -3-3\alpha_{21}^2$.  
The only solution is $\alpha_{21} = 1$, and the
resulting overlattice $\N$ is defined by the matrix
\begin{equation}
\begin{pmatrix}
-12 & 6 \\
6 & -6
\end{pmatrix}\, .
\end{equation}

\vspace*{0.1in}

Now we consider the case of linearly dependent $f_1$ and $f_2$.  In
this case, we have $f_1=x f$ and $f_2 = y f$ for some vector $f\in
\lat$ and $x,y \in \Z$.  The tadpole constraint then gives
$(N_1x^2+N_2y^2) f^2 = -48$.  This is exactly like the one-stack
problem, and solutions for fixed $N_1,N_2$ are easily found.

It is straightforward to implement a computer algorithm that
enumerates all 2-stack solutions, including overlattices.  In Table
\ref{gauge2stack}, we list the gauge groups that are allowed and the
number of vacua with that gauge group, counting all possible amounts
of matter.  We also compute the number of vacua that have distinct
low-energy properties.  There are a total of
574 distinct gauge group and matter content
combinations for low-energy theories.  Table
\ref{mattertwostack} shows the matter hypermultiplets obtained for the
distinct $U(2)\times U(4)\times SO(20)$ models computed using the
formulae listed in Table \ref{tablematter}.

\begin{table}
\centering
\begin{tabular}{||c||c|c|c|c|c|c||}
\hline
\hline
    & 1 & 2    & 3   & 4   & 5 & 6 \\
\hline
\hline
1  &   93 (146) &   $-$  &   $-$  &   $-$  &   $-$  &   $-$       \\      
   \hline 

2  &   54 (67) &   99 (134) &   $-$  &   $-$  &   $-$  &   $-$       \\      
    \hline 
  
3  &    25 (25) &   0 (0) &   3 (3) &   $-$  &   $-$  &    $-$       \\     
     \hline 
  
4   &   27 (27) &   56 (69) &   0 (0) &   18 (18) &   $-$  &   $-$       \\      
    \hline 
  
5   &   9 (9) &   0 (0) &   0 (0) &   0 (0) &   0 (0) &     $-$       \\      
    \hline 
  
6   &   7 (7) &   22 (23) &   0 (0) &   7 (7) &   0 (0) &   5 (5)      \\      
    \hline 
  
7   &   3 (3) &   0 (0) &   0 (0) &   0 (0) &   0 (0) &   0 (0)      \\     
     \hline 
  
8   &   5 (5) &   9 (9) &   0 (0) &   5 (5) &   0 (0) &   0 (0)      \\      
    \hline 
  
9   &   1 (1) &   0 (0) &   0 (0) &   0 (0) &   0 (0) &   0 (0)      \\     
     \hline 
  
10   &   0 (0) &   5 (5) &   0 (0) &   0 (0) &   0 (0) &   0 (0) \\
  \hline
\end{tabular}
\caption{The results of the enumeration of all two-stack models.  The
  entry $x$ in row $P$ and column $Q$ (with $P \geq Q$) implies
  that there are $x$ vacua with gauge group $U(P)\times
  U(Q)\times SO(32-2P-2Q)$ and distinct matter content. 
  The number in parentheses includes the total number of realizations 
  including overlattices.} \label{gauge2stack}
\end{table}

\begin{table}
\centering
\begin{tabular}{|c|c|c|c|c|c|c|}
\hline
$(m_{\scriptscriptstyle 11},m_{\scriptscriptstyle 12},m_{\scriptscriptstyle 22})$ & $\scriptstyle (\mathbf{2},\mathbf{4})_{(1,1)}$ & $\scriptstyle (\mathbf{2},\mathbf{4})_{(-1,1)}$ & $\scriptstyle (1,1)_{(2,0)}$ & $\scriptstyle (1,\mathbf{6})_{(0,2)}$ & $\scriptstyle (\mathbf{2},20)_{(1,0)}$ & $\scriptstyle (\mathbf{4},20)_{(0,1)}$ \\
\hline
$(-  4, k,-  10), \; | k | \leq 5$ &$5-k$& $5 + k$&  6 & 18 & 0 & 3 \\
$(- 8, k,- 8), \; | k | \leq 6$ &  $6-k$& $6 + k$  &  14 & 14 & 2 & 2 \\
$(- 12, k,- 6), \; | k | \leq 7$ &  $7-k$& $7 + k$&  22 & 10  & 4 & 1 \\
$(- 16, k,- 4), \; | k | \leq 8$ &  $8-k$& $8 + k$&  30 & 6  & 6 & 0 \\
\hline
\end{tabular}
\caption{ Matter hypermultiplets for the 56 $U(2)\times U(4)\times
   SO(20)$ models.  The subscript indicates the charges under the two
   $U(1)$'s. $\mathbf{2}$ and $\mathbf{4}$ denote the fundamental
   representation of $SU(2)$ and $SU(4)$ respectively. $20$ denotes the
   vector representation of $SO(20)$. The fourth and fifth columns are the antisymmetric representations of $U(2)$ and $U(4)$ respectively. The hypermultiplet also contains
   the conjugate representation in each case.  There are other ways to
 realize $U(2) \times U(3)$ as a subgroup of the gauge group by
 saturating the tadpole with other brane combinations.
} 
\label{mattertwostack}
\end{table}

\section{``Dial-a-model''}
\label{sec:dialamodel}

In the preceding section we described how the complete set of vacuum
solutions for magnetized branes on K3 can be systematically enumerated.
Such a systematic categorization of models in a particular class can
be useful for performing statistical analysis of a large family of
models and looking for constraints and correlations in the structure
of the corresponding low-energy theories.  Often, however, study of
string compactifications is motivated by the search for models with
specific physical properties.  For example, one may be interested in
restricting attention to models with specific gauge group and matter
content.  For some parts of the landscape, as discussed in Section
\ref{sec:vacuum}, identifying models with particular features can be a
challenging computational problem, even if the mathematical structure
of the vacua of interest is well understood \cite{complexity}.  The
simple theoretical structure afforded by the lattice embedding
theorems for the K3 magnetized brane models we consider here, however,
greatly simplifies the search for models with specific physical
properties.  In fact, for the models we consider here, we can
immediately identify all models with a particular gauge group, with or
without fixing the matter content.  This can be done by simply
imposing certain conditions on the stack sizes $N_a$ and intersection
matrix $m_{ab}$ and proceeding with the enumeration as described in
the previous subsection, subject to the imposed constraints.  For
example, in the previous subsection we identified all 5 models with
gauge group $U(6) \times U(6) \times SO(8)$.

In searching for models with particular structure, however, we may
have only partial information about the model of interest.  For
example, in searching for a standard model-like construction we know
that we want a gauge group that contains the nonabelian subgroup
$SU(3) \times SU(2)$, but we do not know if there are additional
hidden nonabelian gauge groups, which may be broken at experimentally
inaccessible energy scales and associated with as-yet-undiscovered
massive particles.  This suggests that, rather than identifying all
models with a specific complete gauge group, we may wish to identify
all models that contain a certain group $G$ as an {\it subgroup} of
the gauge group.  In most string constructions, many physical features
(such as the number of matter fields in certain representations of
$G$) will depend only on how the subgroup $G$ is physically realized,
and not on what other gauge symmetries may be present in the model.
Posing the question in this way also dramatically reduces the
computational complexity of the problem.  For a given realization of
$G$ (such as in terms of a specific brane geometry), there may be an
exponentially large number of ways in which other branes complete the
gauge group and matter content.  Thus, searching over all models that
contain $G$ may be an exponentially hard problem, while searching for
distinct realizations of the subgroup $G$ may be a problem of only
polynomial complexity.  This is true, for example, in the intersecting
brane model story mentioned previously.  Searching over all possible
models, such as done in \cite{bghlw}, leads to a combinatorial
explosion of models with each possible realization of a specific gauge
subgroup such as $SU(3) \times SU(2)$.  On the other hand, identifying
all possible ways that $SU(3) \times SU(2)$ can be distinctly realized
as part of an intersecting brane model, independent of what other
gauge components and matter fields arise, decreases the difficulty of
the problem
significantly, despite the added complexity of branes with negative
tadpoles and more complicated SUSY conditions.  A complete analysis of
this problem is given in \cite{Rosenhaus-Taylor}, based on the more
general treatment of \cite{Douglas-Taylor}.  (An earlier search over a
smaller range of models based on similar constraints was carried out
in \cite{Cvetic}).

Thus, we may wish to ask, for example, in how many different ways the
gauge group $G = SU(3) \times SU(2)$ can be realized as a subgroup of
the total gauge group in the smooth K3 models
considered here. Basically, each nonabelian component
of the group is associated with a stack with some  particular flux
$f_a$.  These stacks, however, may not completely saturate the tadpole
condition (\ref{eq:tadpole-theory}), as other  branes
may also contribute to the tadpole.
So we are looking in this case for  all configurations with $f_3, f_2
\in\lattice$, satisfying
\begin{equation}
3f_3^2 + 2f_2^2 \geq -48 \,,
\end{equation}
where the lattice generated by $f_1, f_2$ contains no vectors of norm
squared $-2$.
This is very similar to the general enumeration problem analyzed
above, except that we now have an inequality instead of equality in
the tadpole constraint.  
Also, the constraints placed on $f_1, f_2$ by the Freed-Witten anomaly condition are weaker since other vectors can add to the sum $\frac{1}{2}\sum_{a} N_af_a $ that must be an integral lattice vector.
Otherwise, we can proceed in a similar fashion to the above analysis.  For the problem posed here, we should allow $f_a = 0$
for either of the components of the gauge group, since this would
embed the corresponding component in the residual $SO(32- \sum N_a)$
part of the gauge group. Although with such an embedding $SU(3)\times
SU(2)$ is a subgroup of the overall gauge group, we cannot always
break the gauge group to $SU(3)\times SU(2)$ by the Higgs mechanism
since that would require matter in a a particular representation. In
particular, $SO(32- \sum N_a)$ cannot be broken down to $SU(3)\times
SU(2)$ without changing other parts of the gauge group since the only
matter transforming non-trivially under $SO(32- \sum N_a)$ is in the
vector representation and is also charged under one of the $U
(N_a)$'s.

With these constraints, it is possible to analyze all the
ways in which the gauge group of interest, $G = SU(3) \times SU(2)$
can be realized in abelian magnetized brane models on K3.  
When the K3 is nonsingular, or $G$ appears as a subgroup of the  gauge group,
each such
realization is described by an intersection matrix
\begin{equation}
m = \left(\begin{array}{cc} m_{11} & m_{12} \\ m_{12} &
m_{22}\end{array} \right) = \left(\begin{array}{cc} A & B \\ B &
C\end{array} \right) \label{eq:m2b}
\end{equation}
where $A, B \leq 0$, $3 A + 2B \geq -48$, and $AC-B^2 > 0$.  The
discrete redundancy of these models associated with overlattices is
computed just as above.  
If we consider
adding D5 branes, there are more possibilities since $k$ D5 branes
have an $Sp(k)$ worldvolume gauge group.  In a similar way, we can
identify all ways in which any other group $G$ can be realized as a
subgroup of the full gauge group.  

To further ``dial-a-model'' it may be desirable to fix the number of
matter fields in a particular representation of the gauge group.  To
continue with the preceding example, we may wish to constrain the
number of hypermultiplets in the bifundamental representation of $SU(3)\times SU(2)$
 (``quarks'').  Since
the fundamental and antifundamental representations of $SU(2)$ are
identical, we must include hypermultiplets in both the $(N_3,
\bar{N}_2)$ and $(N_3, N_2)$ representations, which from
Table~\ref{t:multiplets} is given by $-4-f_3^2 -f_2^2$.  This number
is thus divisible by 2.  If, for example, we request that this number
of multiplets is 4, then there are only three possibilities: we have
$(f_3^2, f_2^2)= (-8, 0), (-4, - 4),$ or $(0, - 8)$. This narrows the range of possibilities in an enumeration.

A similar analysis can be carried out for any other desired gauge
group and matter content.  
Furthermore, models with any combination of matter
multiplets compatible with Table~\ref{t:multiplets} for some
particular intersection matrix $m$ can be efficiently enumerated.

\section{Conclusions}
\label{sec:conclusions}

In this paper we have analyzed a simple class of string theory
compactifications.  We have shown that lattice embedding theorems can
be used to give a  classification of magnetized brane models
on K3 giving effective 6-dimensional supersymmetric theories of
gravity and gauge theories with a rank 16 gauge group.  
The gauge
group and matter content of these models are encoded in a set of stack
sizes $N_a$ and an intersection matrix $m_{ab}$.   
This gives a clean
characterization of the constraints on 6-dimensional physics from
these models, coming from the dependence of the gauge group and matter
content on $N, m$, and the freedom in these models, arising from the
arbitrary choice of $N, m$ subject to conditions on the associated lattice $\M$: the tadpole constraint, the  condition that $\M$ is an even
negative-semidefinite lattice, and the condition that there be no norm-squared $-2$ vectors.
 For any $N, m$ that give rise to an $\M$ satisfying
these conditions there is a magnetized brane model on a smooth K3 surface realizing
those parameters.  Such a model is generally unique, though in some
cases there is a discrete redundancy arising from overlattices of the
lattice associated with the intersection matrix $m$. 
Understanding the physical significance of the discrete redundancies
possible in this class of models is an interesting problem that we
leave to future work.

In this paper we have focused on theories associated with smooth K3
compactifications.  This condition imposes the constraint  that
the lattice defined by $m$ does not contain vectors of norm-squared 
$-2$.  While most lattices  with such vectors do not seem to
correspond to models with physical spectra, it is possible that in
some cases there are singular K3 compactifications that give sensible
6D physical theories.  We leave further investigation of these models
to further work.

A further constraint on possible low-energy theories arises from the
Freed-Witten anomaly condition in ten dimensions.  In the analysis of
this paper this condition is built into the construction of the
lattice $\M$, which must contain the vector $\frac{1}{2}\sum_{a}
N_af_a$, where the $f_a$'s characterize the gauge group factors
$U(N_a)$.  Unlike the absence of vectors of norm-squared $-2$, this
condition is not transparent from the point of view of the low-energy
spectrum and data $N_a, m_{ab}$.

The analysis presented here gives a simple example of a region of the
landscape where the range of possible models can be neatly classified.
We have shown how models with particular physical features, such as a
desired gauge group (or subgroup of the full gauge group) can be
simply enumerated.  For example, it is possible to find all
ways in which a certain gauge subgroup $SU(N) \times SU(M)$ can
be realized with a fixed number of matter fields in the bifundamental
representation of these groups.

The lattice structure of the second cohomology group of K3 plays a
central role in the classification of models we have developed here.
While it is not clear that such a simple story will hold for a much
broader class of models, it would be interesting to look for analogous
structure in more complex string compactifications.  Compactification
of the models we have considered here on a further 2-torus gives a
class of ${\cal N} = 2$ supersymmetric theories in 4D that may have
interesting properties.  The structure we have found here in the space
of solutions may be helpful in understanding other related
models, such as compactifications on Calabi-Yau manifolds that can be described as
elliptic fibrations over K3, or K3 $\times T^2$ compactifications with
partially or fully broken supersymmetry.

It may be interesting to incorporate half-integral discrete $B$ flux
into the analysis we have done here.  This may, for example, provide
models with odd numbers of matter fields in appropriate
representations of the gauge group, just as in the ``T-dual'' story of
intersecting branes  \cite{reviews}.
This is related to considering bundles without vector structure, which
has been discussed in a related context in
\cite{orientifold-vector}.

In this paper we have restricted attention to abelian fluxes.  Looking
at nonabelian fluxes giving more general vector bundles would lead to
a useful generalization of the models considered here.  Indeed, the
abelian magnetized brane models studied here lie at specific points in
continuous bundle moduli space.  While these points may have some
special features, we have only made this simplification to make the
story simpler; turning on massless moduli will mix the abelian brane
fluxes, leading to a more general nonabelian point in bundle moduli
space.  In each component of moduli space, the norm-squared of the total
cohomology class $(\sum_{a} N_af_a)^2$ is invariant, but one can
smoothly move between configurations with different combinations of
abelian fluxes.  At general points in the moduli space, the nonabelian
instanton configuration will break some of the symmetry, giving a
lower rank gauge group in the effective field theory.  In the
6-dimensional theory, these transitions can be described by a Higgsing
of part of the gauge group.

The gauge group left invariant in the presence of abelian fluxes takes
the form $\prod_{a=1}^K U(N_a) \times SO(2M)$.  The $U(1)$ factors of
the $U(N)$ gauge group factors are anomalous in general and the
corresponding gauge bosons obtain masses by the Stueckelberg
mechanism, though in some cases $U(1)$ factors remain massless.  We
have not analyzed here which $U(1)$ factors remain massless in the
low-energy theory, and have simply written the gauge factors as
$U(N)$, leaving a more detailed analysis of which $U(1)$ factors are
lifted for further investigation.

One of the best understood classes of K3 compactifications like those
we have considered here, are those where the K3 is at an 
orbifold point in its moduli space \cite{Gimon-Polchinski}.  As explicated
in \cite{Berkoozetal}, the models considered in \cite{Gimon-Polchinski}
involve bundles without vector structure.  These models also involve
nonabelian gauge bundles; each D9-brane carries a fraction of a Dirac 
quantum of flux (though Dirac quantization is still obeyed for fields in the 
adjoint and spinor representations). Thus, these models are not
included in the enumeration discussed in this work, but would be
included in a generalization to incorporate general gauge bundles
without vector structure.

One of the principal limitations of  magnetized brane models and the 
T-dual intersecting brane models is that not all moduli
are stabilized.  It would be interesting to apply the methods
developed in this paper to scenarios in which moduli are stabilized
using fluxes on compactification manifolds involving K3, as in
\cite{drs, Tripathy-Trivedi, ak}.  It may be that a systematic
understanding of the embedding of fluxes (with or without D-branes)
into the K3 cohomology could be realized using lattice embeddings in a
way that would shed light on the physics of these compactifications.

Finally, one goal of this work was to understand the extent to which
string theory constrains the space of realizable low-energy
6-dimensional field theories in a simple corner of the landscape.  We
have found that in this class of models  constraints come from
the tadpole cancellation condition and the dependence of the
6-dimensional gauge group and matter content on the parameters $N_a,
m_{ab}$ (shown in Table~\ref{t:multiplets}) that give a topological
characterization of the models.  A broader class of low-energy
theories can be realized when nonabelian bundle structure is allowed,
particularly when small instantons and singularities of the K3 are
allowed to converge, giving rise to large and fairly arbitrary gauge
group structure. It was shown in \cite{aspinwall-morrison} that in the non-perturbative regime, any simple gauge group below a certain rank can be obtained by coalescing $E_8$ instantons at a singularity.  It would be nice to know
whether essentially any  consistent 6D theory can be realized somewhere in the
landscape, perhaps up to some upper bound on the size of the gauge
group and number of matter fields, or if all 6D supersymmetric
theories arising from string theory must have some structure (other
than anomaly cancellation) related to the constraints arising from the
dependence of the 6D theories considered here on $N, m$

\vskip 0.2in {\it Acknowledgements} 
We would like to thank Allan
Adams, Jim Bryan, Miranda Cheng, Keshav Dasgupta, Jacques Distler,
Michael Douglas, Noam Elkies, Tarun Grover, Ken Intriligator, Abhinav Kumar, Tongyan
Lin, Davesh Maulik, John McGreevy, Ilarion Melnikov,
Greg Moore, David Morrison,
Vladimir Rosenhaus, Bogdan Stefanski, Brian Swingle, Cumrun Vafa,
David Vegh, and Edward Witten for
discussions and for comments related to this work.  
We would like to thank the authors of \cite{Ilarion}, especially Ilarion Melnikov, 
for pointing out a factor of two error in an earlier version of this draft, and for
many useful discussions.
We would like to
thank the Clay Mathematics Institute and participants in the 2008 Clay
K3 workshop for helpful discussions, and thank the Banff International
Research Station for support and hospitality while part of this work
was carried out.  We also acknowledge the Tunnel Mountain trail, where
various mysterious aspects of this project were unravelled, and thank
the guru at the top of the mountain (a.k.a. Joe) for his deep insight
into the physics of K3.  This research was supported by the DOE under
contract \#DE-FC02-94ER40818.

\appendix
\section{Lattice basics and the Even, Unimodular Lattice $\lat$} \label{app_lat}

\begin{figure}[b]
\centering
\includegraphics[width=2.5in]{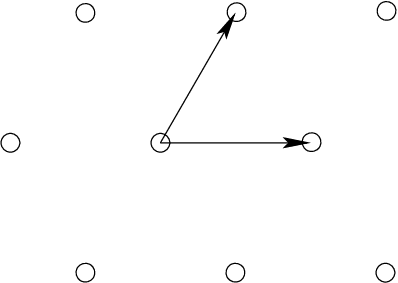}
\caption{Triangular lattice corresponding to the inner product matrix $I^\triangle$.} \label{triangularlattice}
\end{figure}

In this Appendix we review the basic concepts involved in the study of
lattices \cite{Conway-Sloane, serre, nikulin1, morrison}  
 and describe the lattice $\lattice$.  A lattice is
defined as a free $\Z$-module, in other words it is a vector space
defined over the ring of integers $\Z$.  Given a basis
$\{e_1,e_2,\cdots, e_n\}$ we can construct a lattice by taking finite,
$\Z$-linear combinations of the basis elements.  We are interested in
lattices with an even, integral, symmetric, bilinear form (inner
product) denoted by $\cdot\ $.  For any two elements $x,y$ in such a
lattice $\L$, the inner product satisfies $x\cdot y\in \Z$, $x\cdot x \in 2\Z$, $x\cdot
y=y\cdot x$ and is linear in both its arguments.  We can specify the
bilinear form completely by specifying its action on the basis
elements.  Thus, the bilinear form, along with a choice of basis,
defines the inner product matrix $I_{\alpha\beta}:=e_\alpha\cdot
e_\beta$.  For an even lattice, the diagonal elements of the matrix
are all even.  As a simple example of an even lattice, the
lattice defined by the basis elements $\{e_1,e_2\}$ with the inner
product matrix
\begin{equation}
I^{\triangle}=\left(\begin{array}{cc} 2 & 1\\1 & 2\end{array}\right) \label{triangle}
\end{equation}
is the triangular lattice as shown in Figure \ref{triangularlattice}.

A lattice $\L$ is {\it unimodular} if the inner product matrix has
determinant $\pm1$.  The triangular lattice considered above is not
unimodular as it has $\det\ (I^\triangle)=3$.  $\L$ is {\it self-dual} if $\L$ is isomorphic to $\L^*:=\rm{Hom}(\L,\Z)$.  A lattice $\L$ is unimodular $\Leftrightarrow$ $\L$ is self-dual.  The signature of $\L$ is denoted by $(l_+,l_-)$, where $l_+$($l_-$) is the number of positive(negative) eigenvalues of $I$.  A result due to Milnor shows that even, self-dual lattices are very special and exist only when $l_+-l_-\equiv 0(\rm{mod}\ 8)$ \cite{serre}.  

The simplest even, self-dual lattice is the signature (1,1) lattice $U$ given by the inner product matrix
\begin{equation*}
\left(\begin{array}{cc} 0 & 1\\1 & 0\end{array}\right).
\end{equation*}
This lattice makes an appearance in the case of the 4-torus.  The second cohomology group of $T^4$ has the structure of a lattice with inner product given by the wedge product.  $H^2(T^4,\Z)$ is six-dimensional with a basis of two forms given by $\sigma_{ij}:=dx^i\wedge dx^j$, where $x^i$ are coordinates on the $T^4$.  With this inner product, $H^2(T^4,\Z)\cong U\oplus U\oplus U$.

The simplest positive definite, even, self-dual lattice is the root lattice of $E_8$.  In fact, it is the only even, self-dual lattice up to isomorphism in 8 dimensions.  The $E_8$ lattice, which we denote simply by $E_8$, is defined as the integer span of vectors $\{e_1,e_2,\cdots, e_8\}$ with the inner product
\begin{equation}
[e_i\cdot e_j]=\left( \begin{array}{cccccccc}
2 & 0& -1& 0& 0& 0& 0& 0 \\
0 & 2& 0 &-1& 0& 0& 0& 0 \\
-1 & 0& 2&-1& 0& 0& 0& 0 \\
0  & -1& -1& 2& -1& 0& 0 &0 \\
0 & 0 &0 &-1& 2& -1& 0& 0 \\
0 & 0 &0 &0& -1& 2& -1& 0  \\
0 & 0 &0 &0& 0 &-1& 2 &-1 \\
0 & 0 &0 &0& 0 &0 &-1 &2 \\
\end{array} \right) \label{e8abstractbasis}
\end{equation}
This lattice, $E_8$, can be defined in a more physical way as a
discrete subset of $\R^8$.  It is defined as the set of points
$(x_1,\cdots, x_8)\in \R^8$ with $\sum_i x_i \in 2\Z$ such that either
$x_i\in \Z\ \forall \ i$ or $x_i \in \Z+\frac{1}{2} \ \forall \ i$.
This presentation of the $E_8$ lattice is useful in understanding
properties of the lattice that would seem mysterious in the more
abstract definition.  For example, given any positive, even integer
$n$, there is a vector (usually many) of norm-squared $n$ in the $E_8$
lattice.  It is easy to prove this using Lagrange's four square
theorem, which states that any positive integer can be written as a
sum of four integer squares.  We can write
$n=x_1^2+x_2^2+x_3^2+x_4^2$, for $x_i\in \Z$.  Since $n =
(x_1+x_2+x_3+x_4)^2 -2(x_1x_2+\ldots)\in 2\Z$, we have
$x_1+x_2+x_3+x_4\in 2\Z$.  Therefore, the point
$(x_1,x_2,x_3,x_4,0,0,0,0)$ belongs to the $E_8$ lattice and has norm
$n$.  The total number of vectors of any given norm-squared can be computed
using the theta series corresponding to $E_8$ \cite{Conway-Sloane}.
There are numerous vectors of any given norm, for example there are
240 vectors of norm-squared 2, 17520 vectors of norm-squared 8 and 140400 vectors of
norm-squared 16.  Vectors of the same norm-squared can, however, map into one another
under symmetries or relabellings of the lattice.  We are interested in
the number of vectors of fixed norm-squared modulo automorphisms(relabellings)
of the lattice.  The number of vectors up to automorphism of norm-squared $2n$
in $E_8$ is the sequence A008350 in \cite{e8} with elements $\{1, 1,
1, 1, 2, 1, 1, 2, 2, 2, 2, 2,\cdots\}$.  These numbers can be
enumerated using a computer program.  The question of existence and
uniqueness of vectors of given norm-squared has a simple answer in the case of
{\it indefinite} signature, even, self-dual lattices and is explored in
Section \ref{sec:single-embedding} and Appendix \ref{app_embed}.  

The number of euclidean, even, self-dual lattices grows with
dimension.  In 16 dimensions there are two lattices up to isomorphism,
and above 24 dimensions the number of lattices grows rapidly
\cite{Conway-Sloane}.  In the case of indefinite, even, unimodular
lattices, the story is simpler, and there is a complete classification
due to Milnor.  The two lattices $U$, $E_8$ constitute a ``basis'' for
such lattices \cite{serre}.  Every even, unimodular lattice of
signature $(l_+,l_-)$ with $l_+,l_-$ strictly positive is isomorphic
to $U^p\oplus E_8(\pm1)^q$ for $p,q\in \Z$.  The lattice $E_8(\pm1)$
is the $E_8$ lattice with the inner product matrix multiplied by
$\pm1$.  For a K3 surface S, $H^2(S,\Z)$ is an even, self-dual lattice
of signature (3,19) \cite{Asp96a}.  Milnor's classification implies
that $H^2(S,\Z)\cong \lat:=U\oplus U\oplus U\oplus E_8(-1)\oplus
E_8(-1)$.  In the case of the $E_8$ lattice, as discussed above, there
is an alternate representation of the lattice as points in $\R^8$,
which can be more useful in answering certain questions than the
abstract definition in equation (\ref{e8abstractbasis}).  Similarly,
there is a more physical presentation of $\lat$, which can be useful
in answering some questions about K3 surfaces.  This presentation of
$\lat$ is motivated in the context of K3 surfaces and discussed in
further detail in Appendix \ref{app_kummer}.  Here, we simply state
the results of the construction in the form of a basis for $\lat$.
Consider the vectors $\{\pi_{12},\pi_{34},\pi_{31},\pi_{24},\pi_{14},\pi_{23}
\}$ which form a basis for the signature (3,3) even lattice
$U(2)\oplus U(2)\oplus U(2)$ with inner product matrix
\begin{equation}
\left(\begin{array}{cccccc} 0 &\ 2 &\ 0 &\ 0 &\ 0 &\ 0  \\ 2 &\ 0 &\ 0 &\ 0 &\ 0 &\ 0 \\0 &\ 0 &\ 0 &\ 2 &\ 0 &\ 0 \\0 &\ 0 &\ 2 &\ 0 &\ 0 &\ 0 \\ 0 &\ 0 &\ 0 &\ 0 &\ 0 &\ 2\\0 &\ 0 &\ 0 &\ 0 &\ 2 &\ 0 \end{array} \right).
\end{equation}
Here $\pi_{ij}\cdot \pi_{kl}=2\epsilon_{ijkl}$.  Let the vectors
$\{E_0, E_1, \cdots, E_{15}\}$ form a basis of a Cartesian lattice
with inner product $E_i\cdot E_j=-2\delta_{ij}$.  A basis for $\lat$
can be written down in terms of linear combinations of the 
vectors $\{\pi_{12},\pi_{34},\pi_{31},\pi_{24},\pi_{14}, \pi_{23},
E_0, E_1, \cdots, E_{15}\} \in\R^{3, 19}$ and is shown in equation (\ref{basis319}).
The choice of vectors $\pi_{ij}, E_i$ may seem arbitrary, but in fact
these have a geometric interpretation in the construction of a K3
surface by a blow-up of $T^4/\Z_2$.

We conclude this Appendix by defining a general
{\it primitive lattice embedding},
discussed more concisely in \ref{sec:multiple-stack-models}.
A vector $x$
in a lattice $\M$ is a {\it primitive} vector if $x\neq d x'\ \forall
\ x'\in \M, \ 1 <d \in\Z$.  In other words, $x$ is primitive as long as it is
not a (non-trivial) multiple of another vector in the lattice.  For
example, in a lattice generated by $\{e_1,e_2\}$, the vectors $e_1,
e_2, e_1+2e_2$ are primitive, while the vectors $2e_1, 12e_1+18e_2$
are not primitive.  An embedding of $\M$ into $\L$ is specified by an
injective, linear map $\phi:\M \rightarrow \L$ that preserves the
bilinear form.  Such an embedding is said to be {\it primitive} if for
all primitive $x\in \M$, $\phi(x)$ is primitive in $\L$.
Alternatively, the embedding is primitive if the quotient
$\L/\phi(\M)$ is a free $\Z$-module.  These two definitions of
primitive embeddings are equivalent.  For example, consider the
one-dimensional lattice $\M$ spanned by $\{f\}$ which is embedded into
the lattice $\L$ spanned by $\{e_1,e_2\}$ through the map
$\phi(f)=2e_1$.  This embedding is not primitive because the vector
$f\in \M$ is primitive, but it maps to a non-primitive vector $2e_1$
in $\L$.  The quotient $\L/\phi(M)$ is not free and contains the
element $e_1$, which satisfies $2e_1=0$.

\section{Construction of a basis for $\lat$} \label{app_kummer}

In this Appendix we construct an explicit basis for $\lattice$, 
described briefly in Appendix A.  This construction of $\lattice$ is
closely tied to the geometry of the Kummer surface, which is a smooth
resolution of the toroidal orbifold $T^4/\Z_2$ commonly used
in the physics literature.  This basis for $\lattice$ is very useful
for understanding some computations with K3, such as explicit lattice
embeddings. While this basis is referred to frequently in the
literature, we have found that the description in many papers is too
implicit to be immediately useful for computations.  Thus we go into
some detail here in deriving the explicit form of this basis.  The results described in this appendix are primarily
based on the presentations in  \cite{nikulin2, morrison, 
 nahmwendland, abhinavkumar}.

We begin by reviewing the construction of the Kummer surface $X$ as
the blow-up of $T^4/\Z_2$.  We then describe how one can obtain a
basis for the even, unimodular lattice $H^2(X,\Z)$ starting from the
six 2-cycles inherited from the $T^4$ and the sixteen exceptional
2-cycles from the blow-up of the orbifold.  This gives an alternate
presentation of the (3,19) lattice from the usual $U^3\oplus
E_8(-1)^2$ form.  We also define the {\it Kummer lattice}, which has a
nice structure that is easily understood in terms of the geometry of a
hypercube and has a unique, primitive embedding in $\lat$.  Its simple
structure allows one to produce explicit primitive embeddings of even
lattices into the Kummer lattice, and therefore explicit embeddings
into the (3,19) lattice. We consider some examples of such embeddings in 
Appendix \ref{app_embed}.

\subsection{Motivation for the construction from the geometry of Kummer surfaces}

\begin{figure}
\centering
\includegraphics[width=3.5in]{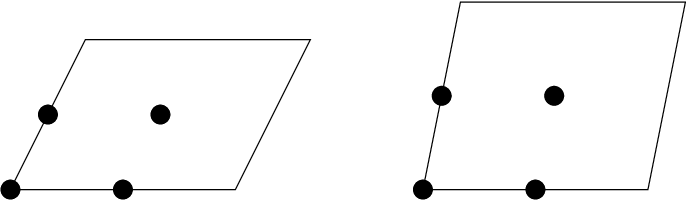}
\caption{$T^4$ represented as a product of two complex tori.  Each 2-torus has four fixed points under inversion of its complex coordinate.  Therefore, the product $T^4$ has sixteen fixed points under the involution.}\label{orbifoldtorus}
\end{figure}

The {\it Kummer surface} is an example of a K3 surface and is
constructed as follows.  Consider a complex two-dimensional torus T
with a $\Z_2$ involution $i$ defined as the inversion of  the two
complex coordinates $z_{1, 2} \rightarrow -z_{1, 2}$.  The quotient manifold $\tilde{T}:=T/\{1,i\}$ is
called the singular Kummer surface of the torus T and is shown in
Figure \ref{orbifoldtorus}.  Each of the sixteen singular points is
locally of the form $\C^2/\Z_2$ and can be blown up by gluing in a
$\P^1$.  Details of this procedure can be found in \cite{Asp96a}.  The
blow-up of all sixteen singularities produces a smooth K3 surface $X$
which is called a Kummer surface.  The blow-up procedure gives sixteen
``exceptional'' 2-cycles (rational curves $\cong \P^1$) which we
denote by $E_0,E_1,\cdots, E_{15}$.  In homology, these sixteen
2-cycles have intersection numbers given by $E_i\cdot
E_j=-2\delta_{ij}$.  In addition, $X$ inherits six 2-cycles from the
$T^4$, which are invariant under the $\Z_2$ orbifold action.  These
2-cycles, denoted by $\pi_{ij}$ are the Poincar\'e duals of the
2-forms $dx^i\wedge dx^j$, where $x^i, i=1,2,3,4$ are the coordinates
on the $T^4$.  In the quotient space $T^4/\Z_2$ the 2-cycles
$\pi_{ij}$, which we sometimes refer to as ``toroidal cycles'', have
an intersection number given by
\begin{equation}
\pi_{ij}\cdot \pi_{kl} = 2\, \epsilon_{ijkl}
\end{equation}
The extra factor of 2 relative to the intersection number of 2-cycles
on $T^4$ is due to the $\Z_2$ quotient which reduces the volume of
space by half thereby increasing the number of intersections by a
factor of 2.  Thus, we have a total of 22 2-cycles (6 toroidal cycles
and 16 exceptional cycles) which are in $H_2(X,\Z)$.  We know that the
lattice $H_2(X,\Z)$ is an even, unimodular lattice, but the lattice
spanned by the 22 cycles we have is not unimodular and in fact has
determinant $-2^{22}$.  We know that $H_2(X,\Z)\cong U^3\oplus
E_8(-1)^2$, but this presentation of the lattice does not make it
clear which 2-cycles come from the torus and which 2-cycles are
exceptional.  In this section, we construct a basis for $\lat$
starting from the toroidal cycles $\pi_{ij}$ and the exceptional
cycles $E_i$.

\subsection{The Kummer lattice} 
\begin{figure}
\centering
\includegraphics[width=4in]{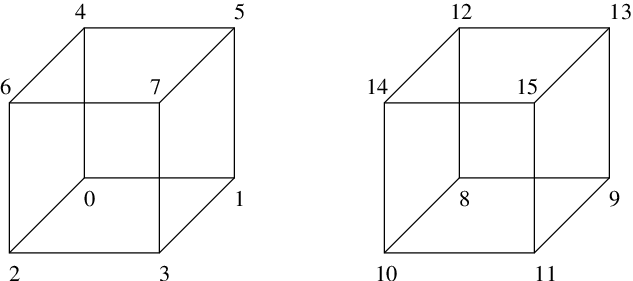}
\caption{The elements of $I$ represented as the vector space $\F^4$ over the field $\F$.  The point $(x_1,x_2,x_3,x_4)\in \F^4$ corresponds to the element $x_1+2x_2+2^2x_3+2^3x_4\in I$.  $\F^4$ is drawn as the two hyperplanes $x_4=0$ and $x_4=1$.} \label{hypercube}
\end{figure}

\begin{figure}
\centering
\includegraphics[width=\textwidth]{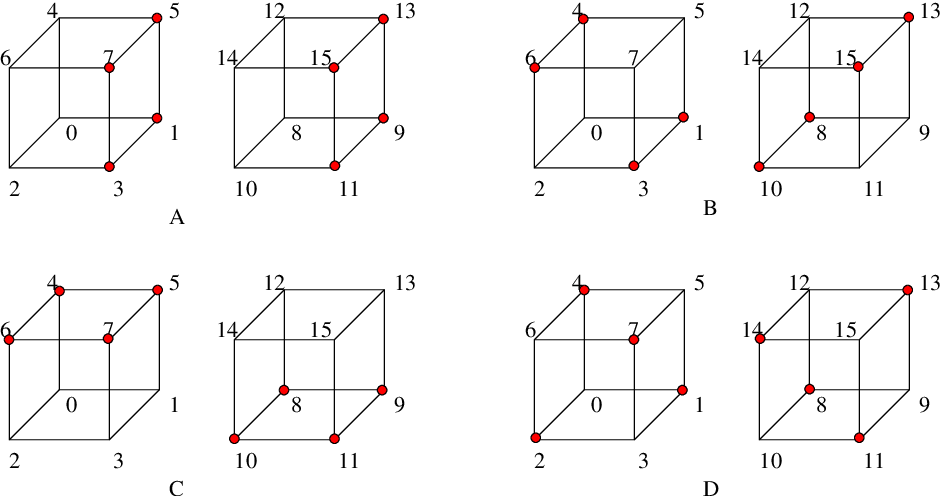}
\caption{Shown above are some hyperplanes in $\F^4 \ - \ A(x_1=1), \ B(x_1+x_3+x_4=1), \ C(x_3+x_4=1),\  D(x_1+x_2+x_3+x_4=1)$.} \label{hyperplanes}
\end{figure}

The starting point of our construction is the Kummer lattice which plays an important role in the study of Kummer surfaces.  This connection was made clear in  \cite{nikulin2}, and most of this section is based on that work.  The Kummer lattice is constructed starting with a set of sixteen orthogonal vectors $E_i, i\in I=\{0,1,\cdots, 15\}$ satisfying $E_i\cdot E_j=-2\delta_{ij}$.  The index set $I=\{0,1,2,\cdots , 15\}$ has the structure of a vector space $\F^4$ over the finite field $\F=\{0,1\}$\footnote[9]{The set $\F=\{0,1\}$ has the structure of a field with addition defined modulo 2 and multiplication defined in the usual way.  The vector space $\F^4$ is the set of 4-bit sequences, where addition and scalar multiplication are carried out bitwise.}.  This vector space can be visualized as the vertices of a four-dimensional hypercube and is drawn in Figure \ref{hypercube}.
A hyperplane in $\F^4$ is defined as a subset of $I$ that satisfies a linear equation of the form $a_1x_1+a_2x_2+a_3x_3+a_4x_4=a_5$, where $a_i\in \F$ and $(x_1,x_2,x_3,x_4)\in \F^4$ are points on the hypercube.  When $(a_1,a_2,a_3,a_4)=(0,0,0,0)$, we have two limiting cases - when $a_5=0$, we have the hyperplane $I$, and when $a_5=1$, we have the null set.  Except for these two cases, every hyperplane has 8 points.  Figure \ref{hyperplanes} shows some examples of hyperplanes in $\F^4$.  Let $Q$ denote the full set of 32 hyperplanes, including the null set and the set $I$.  The Kummer lattice, denoted by $K$, is defined as the set of integer linear combinations of the elements 
\begin{equation}
\{E_i, \ i=0,1,\cdots, 15\} \, \cup \, \{\frac{1}{2}\sum_{i\in M} E_i,\  M\in Q\}
\end{equation}

To discuss the relevance of the Kummer lattice to Kummer surfaces, we first define the Picard lattice of a K3 surface.  It is the group of holomorphic line bundles on a K3 surface, where the group operation is the tensor product.  A line bundle is specified by its first Chern class which is an element of $H^2(X,\Z)$.  A complex structure on $X$ induces a Hodge decomposition $H^2(X,\C):=H^2(X,\Z)\otimes \C = H^{2,0}(X)\oplus H^{1,1}(X)\oplus H^{0,2}(X)$ as discussed in Section \ref{sec:basics}.  Thus, a holomorphic line bundle is specified by an element of $Pic(X):=H^2(X,\Z)\cap H^{1,1}(X)$.  At a general point in the complex structure moduli space, a K3 surface has no holomorphic line bundles.  This is because the  space $H^{1,1}(X)$ is orthogonal to the real 2-plane defined by $\Omega$ (holomorphic 2-form) in $\R^{3,19}$ and in general does not pass through the points of the 22-dimensional integral lattice $H^2(X,\Z)$.  The Kummer surface obtained by the blow-up of $T^4/\Z_2$ is very special since it contains sixteen orthogonal holomorphic curves (equivalent to holomorphic line bundles).  Thus, the Picard lattice of a Kummer surface contains the lattice spanned by $\{E_i,i\in I\}$ as a sub-lattice.  It was shown in  \cite{nikulin2} that in fact, the Picard lattice of a Kummer surface contains the Kummer lattice as a {\it primitive} sub-lattice\footnote[8]{If $X$ is a K3 surface, then $X$ is Kummer $\Leftrightarrow$ The Kummer lattice has a unique, primitive embedding in $Pic(X)$.}

\subsection{Construction of a basis for $\lat$}

We now use the notion of primitive embeddings, defined in Section
\ref{sec:multiple-stack-models} and Appendix \ref{app_lat}, to
construct a basis for $\lat$ starting from the Kummer lattice, which
is a primitive sub-lattice of $\lat$.  This is an example of ``gluing
theory'', which is described in some detail in \cite{Conway-Sloane}.
This approach was used to enumerate 
all the even, unimodular lattices in 24 dimensions.

\begin{figure}
\centering
\includegraphics[width=\textwidth]{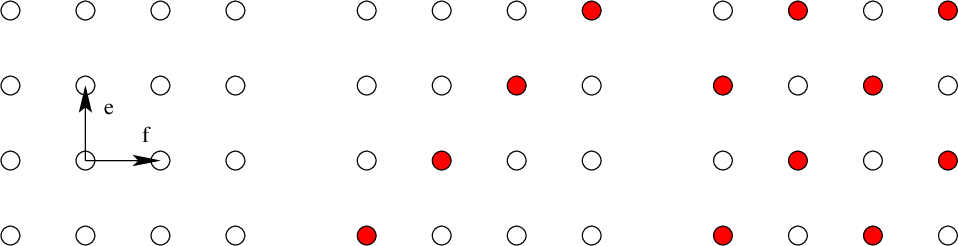}
\caption{The lattice $\L$ is shown on the extreme left.  The diagram in the middle is the lattice $\M$ embedded in $\L$, shown with filled circles.  The third diagram shows the lattice $\L$ as an overlattice of $\M\oplus \M^\perp$ of index 2.} \label{overlattice}
\end{figure}

Before discussing the $\lat$ case, we consider a simpler example which
will serve to illustrate the method.  Consider the lattice $\L:=\{\Z
e+\Z f\}$ with $e^2=f^2=1, e\cdot f=0$ as shown in Figure
\ref{overlattice}.  This is an odd, unimodular lattice.  The lattice
$\M=\Z v$ with $v^2=2$ has a primitive embedding $\phi: \M\rightarrow
\L$ with $\phi(v)=e+f$.  Now, the lattice $\M$ has discriminant $\det
(M) = 2$.  The orthogonal complement $\M^\perp\subset \L$ is the
lattice $N=\M^\perp=\{\Z w\}, \ w=e-f$ with disciminant $\det(N)=2$.
Thus, we have $\det(M)=\det(N)$.  The lattice $\M\oplus \N = \{\Z v+\Z
w\}=\{xe+yf | x,y \in \Z,\ x \equiv y\ ({\rm mod} \ 2)\}$.  The quotient
$\L/(\M\oplus \N)$ is the finite abelian group $\Z_2= \{0, e\}$.
The lattice $\L$ is therefore an overlattice\footnote[3]{As discussed
  in Section  \ref{sec:primitive-over}
a lattice
$\L$ is an overlattice of a lattice $\M$ if the lattice $\M$ has an
embedding into $\L$, and the quotient $\L/\M$ is a finite, non-trivial
abelian group.} of $\M\oplus \N$ with index 2 (See Figure
\ref{overlattice}).  So if we were to construct the unimodular lattice
$\L$ starting from $\M\oplus \N$, we would have to include additional
fractional linear combinations of $v$ and $w$ since $\det(M\oplus N) =
4$.  To determine which precise linear combination we must add, notice
that $(\M\oplus \N) \subset \L \subset (\M\oplus \N)^*$.  The lattice
$(\M\oplus \N)^*=\M^*\oplus \N^*$ is an index 2 overlattice of $\L$.
This leads us to look at the finite abelian group $A=(\M^*\oplus
\N^*)/(\M\oplus \N) = (\M^*/\M)\oplus (\N^*/\N) = \{0,\frac{1}{2}v,
\frac{1}{2}w, \frac{1}{2}(v+w)\} \cong \Z_2\oplus \Z_2$.  The lattice
$\M\oplus \N$ is generated by $\{v,w\}$, and to generate a unimodular
overlattice we must add one or more elements from $A$ to the
generating set.  Since we want the overlattice $\L$ to be integral,
there is a unique choice --- $\frac{1}{2}(v+w)$.  Recall that $v=e+f,
w=e-f \Rightarrow \frac{1}{2}(v+w)=e$.  Adding the vector $e$ to the
generating set $\{v,w\}=\{e+f, e-f\}$ gives the unimodular lattice
$\L$.

Now, we use a similar approach to construct the unimodular lattice
$\lat$ starting from $K$ and its orthogonal complement $K^\perp$,
which we denote $\Pi$.  This construction has been discussed in
\cite{nikulin2, morrison, nahmwendland}.  Since $\lat$
is a unimodular lattice and $K$ is a primitive sublattice with
$\det(K)=2^6$, the orthogonal complement $\Pi=K^\perp$ is an even,
signature (3,3) lattice with $\det(\Pi)=-2^6$ \cite{morrison}.  The
lattice $\Pi$ consists of 2-cycles in the Kummer surface $X$ that are
orthogonal to {\it all} the exceptional cycles $E_i, \ i=0,1,\cdots,
15$.  From the construction of the Kummer surface $X$, it is clear
that these 2-cycles are precisely the cycles that descend from the
$T^4$.  As discussed in Appendix \ref{app_lat}, $H_2(T^4,\Z)\cong
H^2(T^4,\Z)$ is an even, signature (3,3) lattice isomorphic to
$U\oplus U\oplus U$.  If we denote the four 1-cycles of $T^4$ by
$\sigma_i, \ i=1,2,3,4$, the six 2-cycles are given by $\sigma_{ij},
i,j \in \{1,2,3,4\}, \ i < j$.  These 2-cycles satisfy
$\sigma_{12}\cdot \sigma_{34}=1, \ \sigma_{12}\cdot \sigma_{13}=0,
\sigma_{13}\cdot \sigma_{42}=1, \ $ etc.  $X$ is constructed as the
quotient of $T^4$ by the $\Z_2$ involution.  Each 2-cycle
$\sigma_{ij}$ in $H^2(T^4, \Z)$ is invariant under this $\Z_2$ action
and under the image of quotient projection, $\sigma_{ij}\rightarrow
\pi_{ij}$.  The lattice $H^2(X,\Z)$ spanned by the 2-cycles $\pi_{ij}$
in the quotient space is isomorphic to $\Pi:=U(2)\oplus U(2)\oplus
U(2)$.  Therefore, $\Pi = K^\perp \cong U(2)\oplus U(2)\oplus U(2)$
and as expected $\det(\Pi)=-2^6$.  The lattice $\lat$ is therefore an
overlattice of the orthogonal sum of the lattice $\Pi$ (toroidal
cycles) and the Kummer lattice $K$ (exceptional cycles) denoted $\Pi
\oplus K$.

As discussed in the previous example, to generate the even, unimodular overlattice of $\Pi\oplus K$, we must add to the generators of $\Pi\oplus K$ elements from the quotient $A(\Pi) \oplus A(K)$.  Here $A(\M)$ denotes the group $\M^*/\M$ associated with a lattice $\M$.  Both abelian groups $A(\Pi)$ and $A(K)$ have $2^6$ elements since $|\det(\Pi)|=|\det(K)|=2^6$.  Since $\Pi=U(2)^3$, $A(\Pi)=A(U(2))\oplus A(U(2))\oplus A(U(2))$.  It is easy to see that $A(U(2))\cong \Z_2\oplus \Z_2$ and hence $A(\Pi)\cong \Z_2^6$.  $A(\Pi)$ is generated by the set $\{\frac{1}{2}\pi_{12}, \frac{1}{2}\pi_{13},\frac{1}{2}\pi_{14},\frac{1}{2}\pi_{23},\frac{1}{2}\pi_{24},\frac{1}{2}\pi_{34}\}$.  The elements of the quotient $A(K)=K^*/K$ can be written in terms of two-dimensional planes in the $\F^4$ geometry \cite{nahmwendland} as follows -
\begin{equation}
K^*/K = \rm{Span}\{\frac{1}{2}\sum_{i\in P} E_i \ |\ P \subset I \mbox{ is a two-plane.}\} \label{twoplanedef}
\end{equation}
\begin{figure}
\centering
\includegraphics[width=\textwidth]{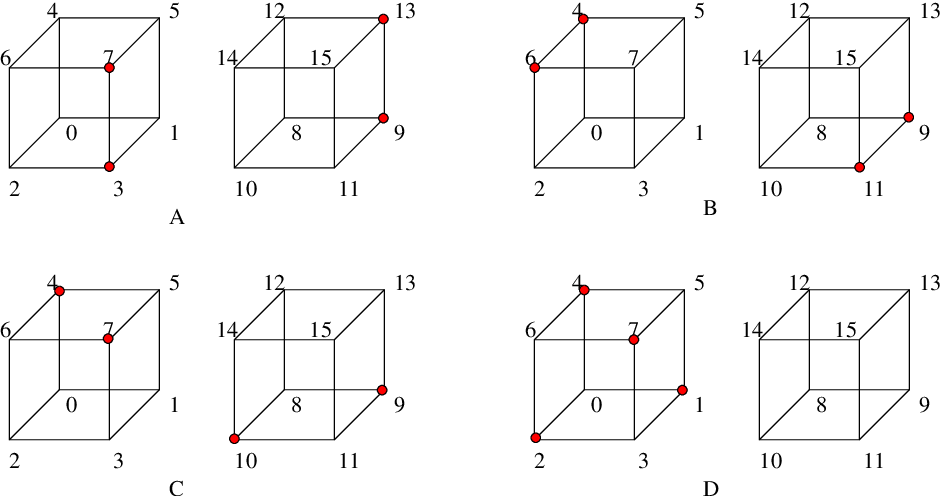}
\caption{2-planes in $\F^4$ constructed by taking intersections of hyperplanes.} \label{twoplanes}
\end{figure}
The set of two-planes in $\F^4$ is obtained by taking all possible
intersections of two hyperplanes.  Some examples of two-planes are
shown in Figure \ref{twoplanes}.  To understand why this is the case,
recall that the elements of $A(K)=K^*/K$ are maps from $K$ to $\Z$ by
the action of the inner product.  Since $E_i^2=-2$, each (non-trivial)
element $x\in A(K)$ is of the form $x=\frac{1}{2}\sum_{i\in P}E_i$ or
else it would map $E_i\in K$ to a non-integer.  Also, the point set
$P$ must not be a hyperplane or $x \in K$ which goes
to zero in the quotient $K^*/K$.  Next, we should make sure that the
elements of $K$ of the form $\frac{1}{2}\sum_{i\in M} E_i$
are mapped to integral values by $x\in A(K)$.  For this to
happen, the point set $P$ must intersect {\it every} hyperplane in an
even number of points.  If $P$ is a 2-plane in $\F^4$, it is
easy to see that this is true.  In addition, we must include linear
combinations (modulo 1 of course) of these elements giving
(\ref{twoplanedef}).  It can be verified that the group $A(K)\cong
\Z_2^6$.

Recall from the toy example  that we cannot add arbitrary elements of $A(K)\oplus A(\Pi)$ to the generating set of $\Pi\oplus K$ since that would make the lattice non-integral.  For example, adding all the generators of $A(K)\oplus A(\Pi)$ would result in the lattice $\Pi^*\oplus K^*$ which has determinant $1/2^6$.  In  \cite{nikulin2} (stated in this form in  \cite{nahmwendland}) it was shown that given a group isomorphism $\gamma: A(\Pi) \rightarrow A(K)$, the lattice $\lat$ can be constructed as the sub-lattice of $\Pi^*\oplus K^*$ as 
\begin{equation}
\lat \cong \{ (x,y) \in K^*\oplus \Pi^* \ | \ \gamma(\bar{x})=\gamma(\bar{y})\} \label{latconst}
\end{equation}
where $\bar{x},\bar{y}$ are the projections of $x,y$ onto the quotient
spaces $K^*/K$ and $\Pi^*/\Pi$ respectively.  In a sense the
isomorphism $\gamma$ ensures that only ``half'' the elements of
$A(K)\oplus A(\Pi)$ are added to the generating set thereby making the
determinant 1.  There is a natural choice for the isomorphism
$\gamma:K^*/K\rightarrow \Pi^*/\Pi$, which is specified by its action
on the generators $\frac{1}{2}\pi_{ij}$ of $A(K)$ as
\begin{equation}
\gamma(\frac{1}{2}\pi_{ij}) = \frac{1}{2}\sum_{i\in P_{ij}} E_i
\end{equation}
Here $P_{ij}$ is the 2-plane in $\F^4$ along the directions $i,j$.  For example, 
\begin{eqnarray*}
\gamma(\frac{1}{2}\pi_{23}) & = & \frac{1}{2}\left(E_0+E_2+E_4+E_6\right) \\ 
\gamma(\frac{1}{2}\pi_{14}) & = & \frac{1}{2}\left(E_0+E_1+E_8+E_9\right)
\end{eqnarray*}

This construction makes it easy to find an explicit basis for $\lat$
in terms of the toroidal cycles $\pi_{ij}$ and the exceptional cycles
$E_i$.  First, we find a basis for the lattices $\Pi$ and $K$.  For
the Kummer lattice a basis can be found by starting with the basis
$\{E_i, i=1,2,\cdots, 16\}$ for a sublattice $K_1\subset K$ and then
looking for lattice points inside the primitive unit cell.  There are
31 such points, given by the ``hyperplane vectors,'' which are not
integer linear combinations of the basis $E_i$.  Choose one such
vector and add it to the basis and drop one of the $E_i$ such that the
new basis spans a lattice $K_2$ with $K_1\subset K_2\subset K$.  This
process is repeated until we have a basis for $K$.  

To find a basis
for $\lat$, start with a generating set 
comprising  (a) The basis vectors $\pi_{ij}$ of $\Pi$, (b)
the basis vectors of the Kummer lattice $K$ obtained by the above
procedure, and (c) extra ``gluing vectors'' of the form
$\frac{1}{2}\pi_{ij}+\frac{1}{2}\sum_{i\in P_{ij}} E_i$, where
$P_{ij}$ is the 2-plane spanned by the directions $i,j$ in $\F^4$.  It
is clear that these vectors generate the unimodular lattice from the
construction of $\lat$ in (\ref{latconst}).  These vectors however are
not linearly independent and form an overcomplete basis of $\lat$.  We
drop vectors from the basis that lie in the span of other vectors in
the basis thereby obtaining a basis for $\lat$ consisting of 22
vectors.  We implemented this procedure and obtained the following
basis (which is not unique in any sense)
\begin{equation}
\begin{array}{c}
 \frac{1}{2} \pi_{12}  +  \frac{1}{2} \left( E_0+E_1+E_2+E_3 \right) \\
 \frac{1}{2} \pi_{13}  +  \frac{1}{2} \left( E_0 +E_1+E_4+E_5 \right) \\
 \frac{1}{2} \pi_{23}  +  \frac{1}{2} \left( E_0+E_2+E_4+E_6 \right) \\
 \frac{1}{2} \pi_{34}  +  \frac{1}{2} \left( E_0+E_4+E_8+E_{12} \right) \\
 \frac{1}{2} \pi_{24}  +  \frac{1}{2} \left( E_0+E_2+E_8+E_{10} \right) \\
 \frac{1}{2} \pi_{14}  +  \frac{1}{2} \left( E_0+E_1+E_8+E_9 \right) \\
 \frac{1}{2} \left( E_0+E_1+E_2+E_3 \right.  +  \left.  E_8+E_9+E_{10}+E_{11} \right) \\
 \frac{1}{2} \left( E_0+E_2+E_4+E_6 \right.  +  \left.  E_8+E_{10}+E_{12}+E_{14} \right) \\
 \frac{1}{2} \left( E_0+E_1+E_4+E_5 \right.  +  \left.  E_8+E_9+E_{12}+E_{13} \right) \\
 \frac{1}{2} \left( E_0+E_1+E_2+E_3 \right.  +  \left.  E_4+E_5+E_6+E_7 \right) \\
 \frac{1}{2} \left( E_8+E_9+E_{10}+E_{11} \right.  +  \left.  E_{12}+E_{13}+E_{14}+E_{15} \right) \\
 E_0,E_1,E_2,E_4,E_5,E_6  ,  E_8,E_9,E_{10},E_{11},E_{12} \end{array} \label{basis319}
\end{equation}
It can be verified that this construction produces an even, unimodular lattice by computing the matrix of inner products.  The signature and determinant can be verified to be (3,19) and -1 respectively.  

\section{Embeddings into Even, Unimodular Lattices} \label{app_embed}

For the construction of magnetized brane models on K3 in this paper, we
are interested in embeddings of even lattices in $\lat$.  
In the case of embeddings into an even, unimodular lattice, the paper
by Nikulin \cite{nikulin1} contains two theorems which will be
extremely useful.  The first theorem, which we quote below, was
originally derived by James \cite{james}.  The second theorem, derived
by Nikulin, strengthens the result by weakening the conditions on the
lattice $\M$.
\begin{theorem}[James] \label{embed1}
Let $\M$ be an even lattice of signature $(t_+,t_-)$ and let $\L$ be an even, unimodular lattice  of signature $(l_+,l_-)$.  If $t_++t_- \leq \min\{l_+,l_-\}-1$, then a primitive embedding of $\M$ into $\L$ exists.  Moreover, this embedding is unique up to an automorphism of $\L$.
\end{theorem}

This theorem is a generalization of the theorem by Wall \cite{wall}
discussed in Section \ref{sec:single-embedding}.  Theorem \ref{embed1}
can be applied to embeddings of $(0,2)$ lattices in $\lat$ in a
similar manner to the embeddings of vectors discussed in Section
\ref{sec:single-embedding}, but breaks down when the rank of $\M$ is
equal to 3 or higher.  Nikulin proved a further extension of this
theorem which applies for higher rank $\M$.
We quote Nikulin's theorem 1.14.4  from 
\cite{nikulin1}, using his remark 1.14.5 to simplify the last
condition, and with an inequality  in the last condition which follows
from the other results in the paper and which is presumably implicit
in Nikulin's statement of the theorem.
\begin{theorem} \label{embed-nikulin}
Let $\M$ be an even lattice of signature $(t_+,t_-)$ and let $\L$ be
an even, unimodular lattice of signature $(l_+,l_-)$.  There exists a
unique primitive embedding of $\M$ into $\L$, provided the following
conditions hold:
\begin{enumerate}
\item $l_+-t_+ > 0$ and $l_--t_->0$.
\item $l_++l_--t_+-t_- \geq 2+l(A(\M)_p) \  \forall \mbox{ primes }p \neq 2$.
\item $l_++l_--t_+-t_- \geq l(A(\M)_2)$ and if equality holds then $A(\M)\cong \Z_2^3\oplus A'$.
\end{enumerate}
\end{theorem}
To understand the statement of this theorem, we must define
$l(A(\M)_p)$.  Given any even lattice $\M$, there is an associated
finite abelian group $A(\M)$ called the {\it dual quotient group}
defined as the quotient $\M^*/\M$.  Here $\M^*:=\mbox{Hom}(\M,\Z)$
denotes the dual lattice of $\M$.  For every prime $p$ the
$p$-component of $A(\M)$ denoted $A(\M)_p$ is defined as the subgroup
of all elements whose order is a power of $p$.  Then, $A(\M) =
\bigoplus_p A(\M)_p$.  $l(A(\M)_p)$ denotes the number of generators
of the $p$-component of $A(\M)$.  For example, consider the lattice
$\M=U(2)$ defined as the lattice generated by $e$ and $f$ with the
inner products $e\cdot e=f\cdot f = 0, \ e\cdot f = 2$.  The basis for
the dual lattice $\M^*$ is composed of the elements $e^*$ and $f^*$
defined as
\begin{equation}
e^*(e)=1, e^*(f)=0 \mbox { and } f^*(e)=0, f^*(f)=1
\end{equation}
An element $v^* \in \M^*$ is naturally associated with an element in
$\mathbb{Q}e_1+\mathbb{Q}e_2$ using the inner product.  For example,
$e^*, f^*$ can be regarded as the elements $\frac{1}{2}f$ and
$\frac{1}{2}e$ respectively.  Thus the dual lattice can be expressed
as the space $U(2)^* \cong \mbox{Span}\{xe+yf\}$, with $x,y \in
\frac{1}{2}\Z$.  The quotient space $A(U(2))=U(2)^*/U(2)$ therefore is the
finite abelian group
$\{0,\frac{1}{2}e,\frac{1}{2}f,\frac{1}{2}e+\frac{1}{2}f\}$ where
addition is $\mbox{mod }1$.  This is the abelian group $\Z_2\oplus
\Z_2$ and is generated by two elements $\frac{1}{2}e$ and
$\frac{1}{2}f$.  In this case, $l(A(U(2))_2)=2$ and $l(A(U(2))_p)=0$ for
$p\neq 2$.  Using Theorem \ref{embed-nikulin}, this means that the
lattice $U(2)$ has a unique, primitive embedding in $\lat$.  The
lattice $\Pi:=U(2)\oplus U(2)\oplus U(2)$ has $l(A(\Pi)_2)=6$ and
$l(A(\Pi)_p)=0$ for $p\neq 2$, and therefore also has a unique,
primitive embedding in $\lat$.

We can derive a simple, but useful upper bound on the number of generators of $A(\M)_p$.  Every finite, abelian group can be decomposed into its $p$-components as $A(\M) = \bigoplus_p A(\M)_p$; there is a natural projection $\pi_p:A(\M) \rightarrow A(\M)_p$ which is onto.  Therefore,  $l(A(\M)_p)$ is certainly less than the total number of generators of the group $A(\M)$.  If the rank of the lattice $\M$ is $d$, the number of generators of $\M^*/\M$ is at most $d \ \Rightarrow \ l(A(\M)_p) \leq d$.  Since $\M$ is a signature $(0,d)$ lattice, Theorem \ref{embed-nikulin} can be used as long as $d \leq 10$.  This provides a proof of Theorem \ref{embed2} in Section \ref{sec:multiple-stack-models}.

One can derive a different upper bound on $l(A(\M)_p)$.  Since $A(\M) = \bigoplus_p A(\M)_p$, $|A(\M)|=\prod_{p}|A(\M)_p|$ where $|A(\M)|$ denotes the cardinality of the group.  Therefore the prime factorization of $|A(\M)|$ determines the cardinality of each $p$-component, but does not determine the group structure.  For example, an abelian group of order 4 is isomorphic to either $\Z_2\oplus\Z_2$ or $\Z_4$ and therefore has 2 or 1 generators respectively.  For an abelian group of order 36, we have four possibilities - $\Z_2\oplus\Z_2\oplus\Z_3\oplus\Z_3,\; \Z_4\oplus\Z_3\oplus\Z_3,\; \Z_2\oplus\Z_2\oplus\Z_9,\; \Z_4\oplus\Z_9$ with 4, 3, 3 and 2 generators respectively.  The general statement is that if $p^k$ is the highest power of $p$ that divides $|A(\M)|$, then the candidate abelian groups for $A(\M)_p$ are in one-one correspondence with the set of partitions of $k$.  
For a fixed cardinality of $|A(\M)_p|=p^k$, the number of generators is maximized when $A(\M)_p \cong \Z_p\oplus\ldots\oplus\Z_p$, with $\max\{l(A(\M)_p)\}=k$.  Therefore, given a lattice $\M$ with inner product matrix $m$, $l(A(\M)_p)\leq k$ where $p^k$ is the highest power of $p$ that divides $|A(\M)|$.  This implies $l(A(\M)_p) \leq \fl{\log_p|A(\M)|}$, where $\fl{x}$ denotes the greatest integer less than or equal to $x$.  Using the relation $|A(\M)| = |\det(m)|$  \cite{Conway-Sloane}, we can derive the upper bound
\begin{equation} \label{bound}
l(A(\M)_p) \leq \fl{\log_p|\det(m)|}\, .
\end{equation}

\end{document}